\documentstyle[emulateapj,psfig]{article}

\lefthead{Safi-Harb et al.}
\righthead{A broadband X-ray Study of the Supernova Remnant 3C~397}

\slugcomment{Accepted for publication in the Astrophysical Journal}

\begin{document}

\def\snei{{\it NEI }}
\def\vsnei{{\it VNEI }}
\def\nei{{\it GNEI }}
\def\vnei{{\it VGNEI }}
\def\shock{{\it PSHOCK }}
\def\vshock{{\it VPSHOCK }}
\def\nshock{{\it NPSHOCK }}
\def\vnshock{{\it VNPSHOCK }}
\def\sedov{{\it SEDOV }}
\def\vsedov{{\it VSEDOV }}
\def\srcut{{\it SRCUT }}

\def\vmekal{{\it VMEKAL }}
\def\mekal{{\it MEKAL }}
\def\power{{\it POWER }}

\def\xr#1{\parindent=0.0cm\hangindent=1cm\hangafter=1\indent#1\par}
\def\ea{\it et al.\rm}
\def\exo{{\sl EXOSAT}}
\def\ein{{\sl EINSTEIN }}
\def\gin{{\sl GINGA}}
\def\rosat{{\sl ROSAT}}
\def\asca{{\sl ASCA}}
\def\chandra{{\sl CHANDRA}}
\def\xmm{{\sl XMM}}
\def\2mass{{\sl 2MASS}}
\def\sest{{\sl SEST}}
\def\as{$^{\prime\prime}$ }
\def\am{$^{\prime}$}
\def\rxte{{\sl RXTE}}
\def\hexte{{\sl HEXTE}}
\def\cps{{ counts s$^{-1}$ }}
\def\sol{$\odot$}
\def\solpyr{{M$_{\odot}$ yr$^{-1}$}}
\def\ergpsec{{ergs s$^{-1}$}}

\title{\bf A broadband X-ray study of the supernova remnant
3C~397}

\author{S. Safi-Harb \altaffilmark{1,2}, R. Petre,
\altaffilmark{2}, K. A. Arnaud \altaffilmark{2}, \\
J. W. Keohane \altaffilmark{3}, \\
K. J. Borkowski \altaffilmark{4}, K. K. Dyer \altaffilmark{4},
  S. P. Reynolds \altaffilmark{4},  \\
\& J. P. Hughes \altaffilmark{5} }
\altaffiltext{1}{NAS/NRC Research Associate;
samar@milkyway.gsfc.nasa.gov}
\altaffiltext{2}{
NASA/Goddard Space Flight Center,
Code 662, Greenbelt, MD 20771}
\altaffiltext{3}{The North Carolina School of Science and Mathematics}
\altaffiltext{4} {North Carolina State University}
\altaffiltext{5} {Rutgers, The State University of New Jersey}

\begin{abstract}

We present a broadband imaging and spectral study of the radio bright
supernova remnant (SNR) 3C~397 with \rosat, \asca, \& \rxte. 
A bright X-ray spot seen in the HRI image hints at the presence
of a pulsar-powered component, and gives this SNR a composite
X-ray morphology.
Combined \rosat \ \& \asca \ imaging show that the remnant is highly
asymmetric, with its X-ray emission peaking at the western lobe.
The hard band images obtained with the \asca \ Gas Imaging Spectrometer
show that much of the hard X-ray emission arises from the western lobe,
associated with the SNR shell; with little hard X-ray emission associated
with the central hot spot.

The spectrum of 3C~397 is heavily absorbed, and dominated by thermal
emission with emission lines evident from Mg, Si, S, Ar and Fe.
Single-component models fail to describe the X-ray spectrum, and at
least two components are required:  a soft component characterized by a
low temperature and a large ionization time-scale, and a hard component
required to account for the Fe-K emission line and characterized by a
much lower ionization time-scale.  We use a set of non-equilibrium
ionization (NEI) models (Borkowski \ea \ in preparation), and find that
the fitted parameters are robust. The temperatures from the soft and
hard components are $\sim$ 0.2 keV and $\sim$ 1.6 keV respectively.  The
corresponding ionization time-scales $n_0 t$ ($n_0$ being the pre-shock
hydrogen density) are $\sim$ 6 $\times $10$^{12}$ cm$^{-3}$ s and
$\sim$ 6 $\times$ 10$^{10}$ cm$^{-3}$ s, respectively. 
The large $n_0 t$ of the soft
component suggests it is approaching ionization equilibrium; thus it
can be fit equally well with a collisional equilibrium ionization
model. 

The spectrum obtained with the Proportional Counter Array (PCA) of
\rxte \ is contaminated by emission from the Galactic ridge, with only
$\sim$ 15\% of the count rate originating from 3C~397 in the 5--15 keV
range.
The PCA spectrum allowed us to confirm the thermal nature of the hard
X-ray emission. A third component originating from a pulsar-driven
component is possible, but the contamination of the source signal
by the Galactic ridge did not allow us to determine its parameters, or
find pulsations from any hidden pulsar.

We discuss the X-ray spectrum in the light of two scenarios:  a young
ejecta-dominated remnant of a core-collapse SN, and a middle-aged SNR
expanding in a dense ISM.  In the first scenario, the hot component
arises from the SNR shell, and the soft component from an
ejecta-dominated component.  3C~397 would be a young SNR (a few
thousand years old), but intermediate in dynamical age between the
young historical shells (like Tycho or Kepler), and those that are well
into the Sedov phase of evolution (like Vela).  In the second scenario,
the soft component represents the blast wave propagating in a dense
medium, and the hard component is associated with hot gas encountering
a fast shock, or arising from thermal conduction.  In this latter
scenario, the SNR would be $\sim$ twice as old, and transitioning into
the radiative phase.
 The current picture we present in this paper is marginally consistent
 with this second scenario, but it can not be excluded.  A spatially resolved
spectroscopic study is needed to resolve the soft and hard components
and differentiate between the two scenarios. Future \chandra \ \& \xmm
\ data will also address the nature of the mysterious central
(radio-quiet) X-ray spot.

\end{abstract}

\section{Introduction}

In the standard scenario of a core collapse explosion of a massive
star, the bulk of the stellar envelope is ejected outward at a high
velocity, shocking the surrounding medium, and forming a shell of
diffuse emission, which is ultimately observed as a supernova remnant
(SNR) shell.  The central core collapses to form a neutron star, which
may be subsequently observed as a pulsar.  SNRs are classified
according to their morphology as shells, plerions, or composites.
While the shells have a shell-like structure, and represent the
majority of SNRs (e.g. the Cygnus loop), the plerions show a centrally
bright morphology in radio and X-rays (like the Crab), with no
evidence of emission from a shell.  The composites have both a shell
plus a centrally bright component.  The X-ray emission from the shell
is in most cases thermal, and results from the shocked swept-up
interstellar medium (ISM), with a probable contribution from
reverse--shocked ejecta in young remnants.  The central X-ray emission
in the plerionic composites is non-thermal, and results from
synchrotron radiation from highly relativistic particles injected by
the pulsar.  As the pulsar wind encounters the surroundings, it gets
shocked and forms the synchrotron nebula, seen as a plerion.

In the thermal composites,  the central emission is thermal, and
arises mostly from thermal emission from the swept up ISM.  Rho \&
Petre (1998) refer to this class as `mixed morphology' SNRs, to
distinguish them from the plerionic composites.

3C~397 (G41.1-0.3)  is one of the brightest Galactic radio SNRs,
 and its  classification remains ambiguous.  In the radio, it is
classified as a shell-type SNR,  based on its
 steep spectral index $\alpha$ = 0.48 and shell-like morphology.
  At 1 GHz, it has a flux density of 22 Jy (Green 1998),
ranking the 5th brightest among $\sim$ 100 remnants.  The distance to
G41.1-0.3 has been estimated as greater than 6.4 kpc on the basis of
neutral hydrogen absorption measurements (Caswell \ea \ 1975).
No absorption is seen at negative velocities, locating the
remnant closer than 12.8 kpc.
An HII region lying $\sim$ 7\am \ west of the SNR
is likely a foreground object,
and was located between 3.6 kpc and 9.3 kpc (Cerosimo \& Magnani 1990).
At a distance of 10 kpc, the linear size of the radio shell would be 7
$\times$ 13 pc$^2$.  High-resolution radio imaging of G41.1-0.3
(Becker, Markert, \& Donahue 1985; Anderson \& Rudnick 1993) indicates
that the remnant brightens towards the Galactic plane, and is highly
asymmetric.  It  has  the appearance of a shell edge-brightened in
parts, and lacks the symmetry seen in the young historical SNRs, such
as Cas~A and Tycho.  3C~397 is slightly polarized with an overall
polarized fraction of only 1.5\%.  Kassim (1989) derives an integrated
spectral index $\alpha$ = 0.4, with a turnover at a frequency less than
100 MHz. 
The discrepancy between the spectral indices derived by Kassim and
Green is most likely due to uncertainties in measuring the
total flux density of 3C~397 at centimeter wavelengths, and is attributed
to confusion with the nearby HII region and the Galactic background.
 Anderson \& Rudnick
(1993) investigate the variations of the spectral index across
the remnant, and find variations of the order of $\delta$$\alpha$
$\sim$ 0.2 ($\alpha$ $\sim$ 0.5--0.7).
 The variations do not coincide with variations in the total
intensity.  They suggest that interactions between the expanding SNR
and inhomogeneities in the surrounding medium play a major role in
determining the spatial variations of the index across the remnant.
Dyer \& Reynolds (hereafter DR~99), while finding a similar magnitude
of spectral index variations, did not confirm Anderson \& Rudnick's
detailed spatial results, suggesting that the variations are due to
image reconstruction problems or other difficulties.

3C~397 has no optical counterpart, and is not observed in the UV
 probably because it lies in the Galactic plane.  An IRAS survey of
Galactic SNRs (Saken, Fesen \& Shull 1992) did not yield a positive
identification of the SNR in the far infrared.

In X-rays, 3C~397 was first detected with the \ein \ Imaging Proportional
Counter (IPC) and High-Resolution Imager (HRI), showing two central
regions of enhancement (Becker \ea \ 1985), neither of which
correlates with any bright radio feature.
The IPC data imply for a thermal model an electron
temperature $kT$ $\leq$ 0.25 keV and N$_H$ $\geq$ 5 $\times$ 10$^{22}$
cm$^{-2}$, implying an optical extinction, $A_v$ $\geq$ 22.5 (Gorenstein 1975).
 The quality of the data was however poor, and clearly other
models were not ruled out.

\rosat \ observations of 3C~397  with the PSPC (Rho 1995, Rho \& Petre
1998, DR~99) reveal 2\farcm5 $\times$ 4\farcm5 diffuse emission, with
central emission and an enhancement along the western edge. Spectrally, thermal
models are favored over non-thermal models, which yield a very steep
power law index (Rho 1995).  The spectrum is generally described by a
thermal plama with $kT$ $\sim$ 1.7 keV. However the fit is poor, probably due to
a combination of abundance effects, nonequilibrium ionization, and the
unrealistic assumption of a single temperature.  3C~397 has similar
characteristics to the `mixed-morphology' SNRs, in that it has a centrally bright X-ray
morphology characterized by a thermal X-ray spectrum; however, the
quality of the \rosat \ data and the narrow bandpass of \rosat \ does
not allow for an accurate determination of its X-ray emission
mechanism.  This left its classification as a `mixed-morphology' candidate highly
uncertain (Rho \& Petre 1998).

Combined \rosat \ HRI and high resolution radio images of 3C~397 (DR~99)
reveal a more complicated morphology.  The SNR appears brightest along
the western edge of the shell, in both the radio and X-ray images.
Moreover, a   bright spot was found with the \rosat \ HRI, at the
center of the SNR shell,
 but not correlated with any radio enhancement (Figure 12 in DR~99).
 No pulsations are found to be associated with the X-ray hot spot.

In this paper, we present a broadband X-ray study
 of 3C~397 with the \rosat, \asca \ and \rxte \ satellites.  While \rosat
\ has the highest spatial resolution, \asca \ has the advantage over
\rosat \ in its broad bandpass and higher spectral resolution, which
allowed us to detect strong emission lines typical of those seen in
the spectra of young SNRs.  \rxte \ complements the other observations
with  its higher energy coverage which allow us to to search for a hard
non-thermal component.  A preliminary analysis of the \asca \ GIS data
 was performed by
Keohane (1998) and Reynolds \& Keohane (1999), in order to obtain an
upper limit on a possible nonthermal component resulting from the
extension of the radio spectrum including an exponential cutoff in the
electron distribution. The authors found that the radio spectrum had to
begin rolling off around $3 \times 10^{16}$ Hz in order to avoid
exceeding the continuum around 1 keV.  Though they did not perform
extensive self-consistent spectral fitting, they found that a very hard
component was also necessary to explain the continuum beyond a few
keV.

The time resolution of \rxte \ ($\sim$ 1 $\mu$s) allowed us to
search for pulsations down to the millisecond range.  The analysis of
the data was however complicated by the contamination of the source
spectrum with  the emission from the Galactic ridge.  The \rxte \ data
have allowed us to:  1) confirm the presence of the hard component
required to fit the \asca \ data and better constrain its parameters;
2) find evidence of a third weak component, whose parameters were
poorly determined due to the contamination by Galactic ridge; and 3)
set an upper limit on the flux from a possibly hidden compact source.

The paper is organized as follows: Section 2 summarizes the
observations.  In Sections 3 \& 4, we present the spatial and spectral
results with \asca\ and \rosat.  In Section 5, we present the \rxte \
results.  In Section 6, we summarize the results from the timing
analysis. We discuss the implications of our results in Section 7.
Finally, we summarize our conclusions.

\section{Observations}

\subsection{ASCA}

3C~397 was observed using \asca \ (Tanaka, Inoue, \& Holt 1994) on 1995
 July 4.  We extracted the data from the HEASARC public database, and
 present the observations acquired with the Gas Imaging Spectrometer
 (GIS), and the Solid State Imaging Spectrometer (SIS).  The \asca \
 detectors are sensitive to X-rays in the 0.4--10 keV range, with a
 spectral resolution at 6~keV of 2\% for the SIS and 8\% for the GIS
 ($\sim$$E^{-1/2}$).  The point spread function of the GIS alone is a
 Gaussian with a full width at half max (FWHM) of 30\as \ (at 6 keV,
 $\sim$$E^{-1/2}$).  The intrinsic spatial broadening of the SIS
 detectors is negligible compared to that of the X-ray telescope
 (XRT).  Therefore the spatial resolution of the detectors is limited
 by the point spread function of the X-ray telescope, which has a
 relatively sharp core (FWHM of 50\as), but broad wings (50\%
 encircled radius of 1\farcm5).

The data were screened using the standard process.
The pointing is at $\alpha$ = 19$^h$ 07$^m$ 43\fs20, $\delta$ =
07\arcdeg~13\arcmin~5\farcs9 (J2000). 
The observations were performed using the standard time resolution: 0.5s
for medium bit rate, and 62.5 millisecond for the high-bit rate.  The
SIS data were acquired in 1-CCD mode, and read out every 4s.  For the
timing analysis, we use the GIS high-bit rate data.  For the spectral
analysis, we use both the SIS and GIS data in high-bit rate mode. 
 We note that Chen \ea \ (1999)
have reported the analysis of the same \asca \ SIS data.
In their paper, the authors use a blank sky field for
background subtraction.  In our
paper, we subtract the background from the same field.  This method is
more appropriate since 3C~397 lies in the Galactic ridge, and removing
any contamination from the ridge is necessary before drawing any
conclusion on the source spectrum.  Furthermore, in addition to the
SIS, we analyze the GIS data which are more appropriate than the SIS
for studying the hard component.

\subsubsection{ROSAT}

In order to better understand the origin of the X-ray emission, we
compare the \asca \ images with the high-spatial resolution
images obtained with \rosat, sensitive in the $\sim$ 0.1--2.0 keV
energy range.  3C~397 was observed with the High-Resolution Imager (HRI)
on several occasions in 1994,
 and with the Position Sensitive Proportional Counter (PSPC) on 1992
 October 28 for $\sim$ 4 ksec.  The \rosat \ data have been analysed
and presented elsewhere (Rho 1995, DR~99).  To generate the \rosat
\ images, we extracted the data from the HEASARC public database.  For
the HRI, we used the longest exposure (52.3 ksec), performed on 1994
October 14. The \rosat \ pointing is at $\alpha$ = 19$^h$ 07$^m$
33\fs60, $\delta$ = +07\arcdeg~08\arcmin~24\arcsec~(J2000).

\subsection{RXTE}

3C~397 was observed on six occasions between 1997 December 3, and 1997
December 8 for an effective exposure of 58.4 ks.
The 1\arcdeg \ FWHM field of view (FOV) of
\rxte \ is pointed at $\alpha$ = 19$^h$ 07$^m$ 34\fs99,
 $\delta$ = 07\arcdeg \ 07\arcmin \ 14\farcs9 (J2000).
The Proportional
Counter Array (PCA), consists of 5 collimated Xenon Proportional
Counter detectors with a total area of 6,500 cm$^2$,  an effective
energy range of 2--60 keV, and an energy resolution of 18\% at 6 keV
 (Jahoda \ea \ 1996).  The \hexte \ (Gruber \ea \ 1996) instrument
consists of two clusters of collimated NaI/CsI phoswich detectors with
an effective area of $\sim$ 800 cm$^2$ and an effective energy range of
15--250 keV.

In this paper, we report the observations with the PCA.  The  \hexte
\ count rates are background dominated, and were not
 included.  For the spectral analysis, we use the data in the
standard-2 mode, which provide spectral information.  For the timing
analysis, we use the Good Xenon modes which provide the highest time
resolution ($\sim$1 $\mu$s).

\section{Spatial analysis}

The \asca \ pointing is $\sim$ 5\am \ offset from the central ``hot
spot'' detected with the \rosat \ HRI.  We generate images of 3C~397
corrected for exposure and vignetting.  We use the routine,
$\it{ascaexpo}$, which calculates the net exposure time per sky pixel
(http://heasarc.gsfc.nasa.gov/docs/asca/abc/).  The total time seen by
each sky pixel on the detector is computed using an instrument map
(generated with $\it{ascaeffmap}$) and the reconstructed aspect.  The
output exposure map is subsequently used to normalize the sky image.
In Figure 1, we show the generated images in the soft (0.5--4.0 keV)
and hard (4--10 keV) energy bands of the GIS.  The effective exposure
is 50.7 ksec (both GIS detectors), and the images are smoothed with a
Gaussian with $\sigma$ = 45\as.  At the harder energies, the emission
shows an elongation nearly perpendicular to the Galactic plane.
In Figure 2, we compare the \rosat \ HRI image (left panel) and the hard
band GIS image (right panel).  The brightest features in the \rosat \
image, seen at $\sim$ 1\am--2\am \ east and west of the central hot
spot, correlate with enhancements in the radio shell (DR~99).  It is
clear that these small scale features seen with the HRI cannot be
resolved by the GIS.  However, the overall morphology of the GIS image
shows an elongation along the axis joining the hot spot with the
bright radio edges. The hard emission peaks at the
western lobe, and is not centrally peaked at the 
 HRI hot spot (denoted by a cross),
as would be expected from a plerionic composite.
In Figure 3, we overlay contours from the SIS
hard band on the \rosat \ PSPC image. The SIS contours are correlated
with the PSPC intensity map, and again suggest that the hard emission 
peaks at the western lobe, with some fainter emission associated with
the central spot and the eastern lobe.
While the GIS is more sensitive than the SIS to the hard
X-ray band ($E$ $\geq$ 4 keV), the SIS has a higher efficiency at the
softer energies.  We subsequently examine the softness ratio, defined
as $\frac{0.5-2 (keV)}{2-4 (keV)}$, with the SIS. In Figure 4, we show
the resulting image (left panel), with the total soft band (0.5--4
keV) image obtained with the SIS (right panel).  The HRI contours are
overlayed on both images, to show that there is an enhancement of soft
emission from the central region.

\section{Spectral analysis: \asca}

To study the X-ray spectrum, we have extracted events from a circular
region of radius $\sim$~7\am \ from the GIS field, encompassing the
entire SNR.  For the SIS which has a smaller FOV than the GIS,
 3C~397 fills a large fraction
of the chip, and extends to the edges of the CCD along the
direction perpendicular to the Galactic plane.
We have extracted the source
events from a circular region of radius $\sim$~4.2\am \ to
avoid the CCD chip boundaries.
Since 3C~397 lies $\sim$ 0.3$^\circ$ below the Galactic
plane, and only 41\arcdeg \ in longitude from the Galactic center, the
source spectrum is expected to be contaminated by emission from the Galactic
ridge.  The large FOV of the GIS allowed us to extract a background
spectrum from the same field.
For the SIS, it was also possible to extract a background (of radius $\leq$ 1\am)
from the same chip, since 3C~397 does not fill the FOV along the direction parallel
to the Galactic pane.
This method of background subtraction has the advantage of providing the
most accurate model of any spatial or temporal contamination which
would affect the spectral analysis.
While the extracted SIS spectrum contains most of the emission from the SNR,
it is possible that we are missing part of the source flux.
This was taken into account by introducing a relative normalization between the
SIS and the GIS spectra.  Throughout the paper, we perform the flux
estimates from the different components of the SNR using the GIS.  The
background subtracted count rates in the 0.6--9 keV energy range are
0.524 $\pm$ 0.005 \cps and 0.613 $\pm$ 0.005 \cps from GIS2 and GIS3
respectively.  The corresponding SIS0 and SIS1 count rates are 0.768
$\pm$ 0.007 \cps and 0.580 $\pm$ 0.007 \cps respectively.

In the following, we present our spectral results for both the SIS and
GIS data. 
 We fit the SIS data in the 0.6--9 keV range, and the GIS over
the 0.8--9 keV band. We disregard energies above 9 keV due to the poor
signal-to-noise ratio.  We combine the SIS and GIS data to show a
joint fit, to which we subsequently add the \rxte \ PCA data.

\subsection{Single-component models}

The SIS and GIS spectra are clearly dominated by emission lines, with
strong emission from Mg, Si, S, and Ar; the most prominent feature is
the Fe-K line. We have first attempted to fit the spectra with
Raymond-Smith (RS, Raymond \& Smith 1977) and \mekal (Mewe,
Gronenschild, \& van den Oord 1985; Liedahl \ea \ 1990) models, which
are appropriate for modeling plasma in collisional equilibrium ionization.
A single-component collisional equilibrium ionization
model with solar abundances does not yield an acceptable fit (reduced
chi-squared $\chi_{\nu}^2$ = 6.7, $\nu$ = 735, $\nu$ being the
number of degrees of freedom). 

We subsequently used Sedov models (Hamilton, Sarazin, \& Chevalier
1983, hereafter HSC 1983, Borkowski \ea, in preparation) which follow
the time-dependent ionization of the plasma in a supernova remnant
evolving according to Sedov self-similar dynamics.  These models are
especially important for describing the emission from SNRs whose age
is smaller than the time required to reach ionization equilibrium.
They are a subclass of non-equilibrium ionization (NEI) models, and
include the range of temperatures found in a Sedov remnant.  They can
therefore account for the hotter X-ray emission expected to originate
from the inner parts of such SNRs.  The modifications introduced by
Borkowski \ea \ (in preparation) include improved atomic data (in
particular, Fe L-shell line data are based on theoretical calculations
by Liedahl, Osterheld, \&\ Goldstein 1995) and the possibility of
describing plasmas without electron-ion equipartition, including an
incomplete heating of electrons at a blast wave.  The models are
characterized by three parameters: the shock temperature $T_s$, the
post-shock electron temperature $T_e$ ($\le T_s$), and $\eta$ = $n_0^2
E$, which characterizes the rate at which the plasma relaxes to
ionization equilibrium ($n_0$ is the hydrogen number density in the
unshocked ambient medium, and $E$ is the explosion energy).  An
equivalent parameter to $\eta$ is the ionization time-scale $n_0 t$ =
1.24 $\times$ 10$^{11}$ $\eta_{51}^{1/3}$ $T_{s,7}^{-5/6}$ (cm$^{-3}$
s), where $\eta_{51}$ is $\eta$ in units of 10$^{51}$ erg cm$^{-6}$,
and $T_{s,7}$ is the shock temperature in units of 10$^7$ K.  We find
that the Sedov model provides a better fit, but only accounts for the
X-ray emission up to about 4 keV. Even a Sedov fit with
non-equipartition ($T_e$ $\neq$ $T_i$; where $T_e$ and $T_i$ represent
the electron and ion temperatures respectively) does not provide a
satisfactory fit to the hard component.  In Figure 5, we show the
single-component Sedov fit, characterized by an interstellar
absorption $N_H$ = 2.75 $\times$ 10$^{22}$ cm$^{-3}$, a temperature
$kT_s$ = 0.15 keV, and an ionization parameter of $n_0t$ = 1.86
$\times$ 10$^{12}$ s cm$^{-3}$ ($\chi^2_{\nu}$ = 2.89, $\nu$ = 734).

Varying the metal abundances within the \mekal or the Sedov models improves the
fits. However, the models still do not account for the hard X-ray emission
(above 4 keV), and the fits are unacceptable ($\chi^2_{\nu}$
$\geq$ 2.6).

\subsection{Two-component models}

While the single-component models account for the X-ray emission at
energies below about 4 keV, they fail to fit the higher energy
component (Figure 5).  This result was also found by Chen \ea \
(1999).  We note that even though the Sedov model does include
higher-temperature gas from the remnant interior, this model does not
account for the hard component.  In particular, the most prominent
emission line near 6.55 keV is not accounted for, indicating that the
Fe-K line region cannot be explained by the same component responsible
for the Fe-L emission, and must be characterized by different
ionization parameter values. This result was also found in the young
ejecta-dominated SNRs such as Tycho 
(Hwang, Hughes, \& Petre 1998), Cas~A (Borkowski
\ea \ 1996) and Kepler (Tsunemi, Kinugasa, \& Ohno 1996).

In order to characterize the hard component, we first fit the data in
the 4-9 keV range with a thermal bremsstrahlung (TB) model plus a
Gaussian to account for the Fe-K emission line.  The centroid of the
Fe-K line is at $E$ = 6.55 keV (6.51--6.60, 3$\sigma$), and the TB
temperature is $kT_h$ = 2.5 keV (1.6--4.2, 3$\sigma$).  For a
collisional equilibrium ionization model, such as Raymond-Smith, the
centroid of the strongest Fe-K lines should be $\sim$ 6.7 keV (He-like) and
$\sim$ 6.95 keV (H-like).  The low fitted centroid energy
indicates that the hard component has not
reached ionization equilibrium (Borkowski \& Szymkowiak 1997), and
should be characterized by a NEI model.

We use a number of NEI models (Borkowski \ea, in preparation),
which are now released in XSPEC~11:
\begin{itemize}
\item{\shock, which comprises a superposition of components of
different ionization ages appropriate for a plane-parallel shock. This
model is characterized by the constant electron temperature, $T_e$, and
the shock ionization age, $n_0 t$ (where $n_0$ is the pre-shock
density, and $t$ is the age of the shock; the post-shock density is
constant).}
\item{\snei, a constant-temperature, single-ionization
time-scale NEI model.}
\item{\sedov,  a NEI model based on the Sedov
dynamics, which includes a range of temperatures, as described above. }
\end{itemize}

A proper definition of ionization time-scale is the product of
postshock electron density $n_e$ and age $t$, because the plasma
ionization state depends on $\int n_e dt$. This is the parameter which
enters any NEI model, including the models just mentioned. But the
quantity of most interest here is $n_0t$, which is equal to $n_et/4.8$
for cosmic abundance plasma and the strong shock Rankine-Hugoniot jump
conditions; $n_0$ here includes only Hydrogen. For convenience, we
refer to both $n_et$ and $n_0t$ as the ionization time-scale
throughout this work.

In the 4--9 keV energy range, \asca \ data are fitted equally well with
\snei and \shock models with the solar Fe abundance. With the \shock
model, we obtain  $kT_e$ = 2.45 (1.8--4.2) keV,  $n_0t$ = 3.1
(1.5--8.3) $\times$10$^{10}$ cm$^{-3}$ s, and $\chi_{\nu}^2$ = 1.0
($\nu$=145). As expected, the electron temperature in the \shock model
is equal to the TB temperature, and the plasma is underionized.

We then fit the entire energy range (0.6--9 keV) with various
two-component models, using current NEI models mentioned above, and we
show the results in Table~1.  In fitting the data, we added a Gaussian
near 3.1 keV to account for the emission line from Argon, since current
NEI models do not include emission from this element.  Since the soft
component is characterized by a relatively long ionization time-scale
($n_0t$ $\sim$ 10$^{12}$ cm$^{-3}$, previous section), we can 
represent it using a \mekal model. We also represented the soft
component with the \sedov \ model, but because these fits are
significantly worse (with a reduced $\chi^2 > 2$) we do not include
them in Table~1.  The harder component is characterized by a lower
ionization time-scale, and it is necessary to fit it with a NEI model.
From Table~1, we find that independently of the NEI model used, the
fitted temperatures, ionization time-scales, and emission measures
($EM$) are in reasonable agreement.  The best fit two-component NEI
model, \shock+\shock, yields $N_H$ = 3.21
$\times$ 10$^{22}$ cm$^{-2}$, $kT_l$ = 0.19 keV, $kT_h$ = 1.52 keV,
$n_0t_l$ =5.6 $\times$ 10$^{12}$ cm$^{-3}$ s, and $n_0t_h$ = 6.0 $\times$
10$^{10}$ cm$^{-3}$ s; where $l$ and $h$ refer to the low-temperature
and high-temperature components respectively.  The fit yields a
reduced $\chi^2$ of 1.37 (for 729 degrees of freedom). 
In Figure 6,
we show the \asca \ data fitted with the corresponding model.

We note that when fitting the hard component with a $\it{SEDOV}$ model
(with the soft component fitted with \mekal or \shock), we use both
equipartition ($T_e$=$T_i$) and non-equipartition ($T_e$$\neq$$T_i$)
models.  We find that the non-equipartition model improves the fit,
and the corresponding parameters are: $kT_s$=1.73 keV, $kT_e$=0.86
keV, $n_0t$=9.6$\times$10$^{10}$ (cm$^{-3}$ s).  In Figure 7, we
show the confidence levels for the electron
temperature, $T_e$, versus the shock temperature, $T_s$. For shock
speeds and ionization timescales derived for the hot component,
Coulomb heating is effective, and the mean electron temperature in the
shocked gas is much larger than the postshock electron temperature
$T_e$, and equal to about 1.5 keV.

The hot component temperature in all two-component fits is lower than
the temperature of 2.45 keV derived from fitting the hard (4--9 keV)
component independently (using \shock with solar abundances). In
addition, two-component models always produce too few counts at high
energies. We believe that this is caused by presence of
multi-temperature plasma in 3C~397, with temperatures in the range 0.17
keV -- 2.5 keV, a likely possibility in view of the complex morphology
of 3C~397. Because two-component fits are apparently too simple to
describe \asca \ spectra, we attempted to fit a three-component
model. This has not resulted in a better fit, presumably because the
description of multi-temperature plasma in terms of just 3 components
might still be grossly inadequate, while giving us too many parameters
to be reliably determined from the spatially-integrated X-ray spectrum
alone. More complex multi-component models might give a better fit,
but the problem with a large number of parameters remains, so that we
did not pursue multi-component fitting beyond a 3-component model.

\subsection{Abundances}

Varying the metal abundances improves the spectral fitting, indicating
that at least part of the X-ray spectrum may be associated with an
ejecta component.  Using the \mekal+\shock model, we froze the
abundances of the hard component at their solar values, and varied the
abundances of the elements producing strong lines in the soft
component.  We find that the absolute value of the abundances of the
individual elements are highly uncertain, while their relative values
are nearly the same.  Therefore, we indicate their ratio relative to
Si, relative to solar (given by Anders and Grevesse 1989).
Using the \vmekal model (\mekal with variable abundances) for
the soft component, we allow for the abundances of O, Ne, Mg, Si, S,
Fe and Ni to vary, with Fe and Ni tied together. We find that the fit
improves by a $\Delta\chi^2$ = 324, and yields $\frac{O}{Si}$=4.0
(3.0--5.8), $\frac{Ne}{Si}$=5.8 (2.6--9.7), $\frac{Mg}{Si}$=1.3
(0.9--1.9), $\frac{S}{Si}$=22 (17--28), $\frac{Fe}{Si}$=1.7 (0--4.3)
(2$\sigma$, $\chi^2_{\nu}$=1.16, $\nu$=728).
The apparent high S abundance might be an artifact
of the models used, because the S line is in the energy range where
X-ray spectra from the low- and high-temperature components overlap.

We also investigated the abundances of the hard component, by fitting
the soft component with a \mekal model (with solar abundances), and
using \vshock for the hard component (\shock with variable
abundances).  We fix H, He, C, N, and O to solar; we tie Ni to Fe; and
allow for Mg, Si, S, and Fe to vary.  We find that the fit improves by
$\Delta\chi^2$ = 280, and yields $kT_h$ = 1.39 keV, Mg=2.5\sol, Si=0,
S=1.33\sol, Fe=Ni=1.49\sol.  While varying the
abundances in this way does improve the fit, the inference of strong
Mg and Fe, but no Si, suggests that other possibilities for explaining
the less-than-ideal fit, such as the presence of multi-temperature
plasma, should be examined. We already know that the hard component
temperature is lower in the two-component fits than from fits
to high energy ($> 4$ keV) data alone. An underestimate of the
temperature of the hard component would underestimate the continuum,
which would in turn artificially boost the abundances from this
component.

We also tied the abundances of the soft and hard component
and allowed them to vary.
We used \vmekal  and \vshock  to represent
the soft and hard component respectively.
We fix H, He, C, and N to solar; we tie Ni to Fe; and allow
for  O, Ne, Mg, Si, S, and Fe to vary.
The fit yields a $\chi^2_{\nu}$= 1.16 ($\nu$=728)
with the following abundance ratios:
$\frac{O}{Si}$ = 3.2 (2.6--4.1),
$\frac{Ne}{Si}$ = 2.3 (0.05--3.1),
$\frac{Mg}{Si}$ = 1.2 (0.8--1.4),
$\frac{S}{Si}$ = 2.9 (2.5--3.3),
$\frac{Fe}{Si}$ = 1.8 (1.4--2.3);
the ranges are at the 90\% confidence level.

In view of a possible presence of a low-temperature ejecta component in
3C~397, one might inquire about the type of the supernova (SN) progenitor, by
examining in more detail the abundances determined from fitting the
X-ray spectrum. Numerical models for the
nucleosynthetic yield as a function of the progenitor's mass have been
calculated by Tsujimoto \ea \ (1995) and many others.  The models
predict approximately solar abundances of O and Ne with respect to Si
for core-collapse SNe.  For type Ia explosions, the calculations of
Nomoto \ea \ (1984) indicate negligible O, Ne, and Mg abundances (with
respect to Si), and a large Fe to Si ratio.  The high interstellar
absorption towards 3C~397 did not allow a direct detection of the O and
Ne lines in the $\sim$ 0.7--1 keV range.  However, since these
elements also provide recombination continuum emission, a large O (and
Ne) to Si ratio was shown to improve the fit to the soft component.
Since the O, Ne, and Fe abundances are a
good indicator for the type of the SN explosion, we have tested for
their ratio relative to Si (relative to solar) by:
\begin{itemize}
\item{ Allowing O, Ne, Mg, and Si to vary independently, using
the \vshock  model. 
The $\chi^2$ value decreases from 1,053 to
1,003 ($\nu$=729) and yields the
following ratios:
$\frac{O}{Si}$ = 2.8, $\frac{Ne}{Si}$ = 2.8, $\frac{Mg}{Si}$ = 2.2.}
\item{Setting O to zero, a value consistent with a type Ia
yield, keeping the Fe abundance frozen to solar.
The fit does not improve ($\chi^2$=1,078, $\nu$=729), and
yields $\frac{Ne}{Si}$ = 1, $\frac{Mg}{Si}$ = 2.}
\item{Setting O to zero and allowing Fe to vary. The fit requires no Fe
($\frac{Fe}{Si}\leq$1.3, 2$\sigma$),
with $\frac{Ne}{Si}$ = 1.5,
$\frac{Mg}{Si}$ = 1.1; and a $\chi^2$=1008 ($\nu$=728).}
\item{Forcing a large Fe (Fe $\sim$ 100\sol) as expected
from a type Ia yield, and allowing
O, Ne, Mg, and Si to vary.  The fit gives a $\chi^2$ = 1058 ($\nu$=72=
9), and necessitates very large O, Ne, and Mg abundances relative
to Si.}
\item{Finally, allowing O, Ne, Mg, Si, S, and Fe to vary,
and tying Mg, Si, and S, we find $\frac{O}{Si}$ = 1.95,
$\frac{Ne}{Si}$ = 1.67,  and $\frac{Fe}{Si}$ = 0 ($\le$0.6, 2$\sigma$);
$\chi^2$ = 1025, $\nu$=728. }
\end{itemize}

We conclude that the abundance ratios with respect to Si are certainly
inconsistent with a type Ia yield, but they are consistent with an
explosion of a massive progenitor.  A large S/Si ratio, $\sim$ 4.4,
obtained by varying the S abundance, is inconsistent with both SN
types.  We already mentioned that the large S abundance might be an
artifact of the models used. Although the inferred overabundances of
heavy elements are substantial (larger than in the well-known SNR
Cas~A), the evidence for enrichment of heavy elements is of
circumstantial nature, and a more definite conclusion about the SN
progenitor will probably require the kind of spatially resolved
spectral data that the new generation of X-ray telescopes will
provide.

\section{\rxte \ results}

\subsection{Background}

The \rxte \ PCA instrumental background consists of internal
background as well as the background due to cosmic ray flux and
charged particle events.  We use the latest background model developed
for the analysis of faint sources with the PCA.  This model (L7/240)
accounts for activation in the PCA
(http://lheawww.gsfc.nasa.gov/$\sim$stark/pca/pcabackest.html).  In
Table~2, we list the background-subtracted count rates in the 2.5--20
keV range for the various observation intervals.  We use the XTE/PCA
internal background estimator script $\it{pcabackest \ v2.0c}$ in
order to estimate the PCA background.  We disregard energies below 5
keV, in order to avoid the instrumental Xenon-L edge seen in the
4.5--5 keV range.  To avoid uncertainties in the background
subtraction at the higher energies and to maximize the signal-to-noise
ratio, we analyze the PCA data up to 15 keV only.

In addition to the instrumental background, we have to account for the
emission from the Galactic ridge.  Following Valinia \& Marshall
(1998, hereafter VM~98), we approximate the ridge emission with a
two-component model consisting of a power law with a photon index,
$\Gamma_{GR}$, plus a Raymond-Smith thermal plasma with a
temperature, kT$_{GR}$.  We allow these parameters to vary within the
range determined by VM~98.  In order to determine the normalization, we
examine the scans of the Galactic ridge for a latitude of
-0.25\arcdeg, and within longitudes 30\arcdeg \ and 50\arcdeg \
(after removing the bright sources). We also examine the background
region selected from the \asca \ GIS field of view, and fit its
spectrum combined with the \rxte \ spectrum using the two-component
model described above.  The corresponding flux is subsequently used in
modeling the overall spectrum of 3C~397.  We find that only $\sim$15\%
of the total PCA count rate originates from 3C~397 in the 5--15 keV
range. The flux is dominated by the emission from the
ridge, since \rxte \ has a large FOV and lacks the spatial resolution
needed to resolve the emission from 3C~397.

\subsection{3C~397}

In fitting 3C~397, we freeze the power law index and the RS
temperature of the Galactic ridge, and allow its normalization to span
the 3$\sigma$ range determined with the method described above.  The
spectral fitting with the PCA is insensitive to the soft component
(which dominates up to $\sim$ 2 keV), therefore we represent it by the
\mekal \ model.  For the hard component, we use the \shock model and
find that its parameters are consistent with the \asca \ fit. We find
that a broad Gaussian line is needed to account for the Fe-line
feature seen in the PCA spectrum. It is possible that this line is
associated with the background in the field of 3C~397. The lack of
spatial resolution, and the uncertainties in the ridge model, leave
its origin uncertain.  The \shock model describing 3C~397 hard
component yields  $kT_s$= 1.50 (1.46--1.54, 3$\sigma$) keV,
and  $n_0t$= 7.1 (5.3--9.6)$\times$10$^{10}$
(cm$^{-3}$ s). The corresponding observed flux from 3C~397 $F_x$(5--15
keV) = 3.22 $\times$ 10$^{-12}$ erg cm$^{-2}$ s$^{-1}$; which
corresponds to a luminosity $L_x$(5--15 keV) = 4.0 $\times$ 10$^{34}$
$D_{10}^2$ erg s$^{-1}$.  The fit yields a $\chi_{\nu}^2$=1.68
($\nu$=769).  In Table~3, we summarize the results of this fit, and in
Figure 8 we show the corresponding fit to the combined SIS, GIS, \&
\rxte \ spectra, using the \mekal+\shock model in the 0.6--15 keV
band, and allowing the relative normalization to be a free parameter.

We have further tested for the parameters of the hard component,
independently of the soft component, by fitting the \asca \ hard band
only (4--9 keV) and the PCA spectrum (5--15 keV band), using a \shock
\ model. A broad Gaussian line was again needed to account for the
Fe-line seen in the PCA spectrum.  The \shock model describing the
hard emission from 3C~397 yields  $kT_s$ = 2.3 (2.1--3.1) keV,
 and $n_0t$ = 3.1 (1.5--6.3) $\times$10$^{10}$ (cm$^{-3}$ s),
 with a $\chi^2_{\nu}$=1.08 ($\nu$=170).
 These parameters are consistent with the fit to the
\asca \ hard (4--9 keV) band described previously, and better
constrained.

\subsection{Non-thermal emission?}

The emission from 3C~397 is more complicated than can be simply
described by a two-component thermal model, partly because the emission
from the central region appears at both energy bands. This might be
an indication of the presence of an additional unresolved component,
possibly non-thermal, hinting at the presence of a plerion.  In
addition, the GIS image in the hard energy band (4--9 keV) shows that
the overall morphology of the remnant follows the radio and X-ray
enhancements seen in the radio and HRI image (see Figure 2), indicating
that part of the flux could be non-thermal and associated with highly
energetic electrons accelerated at the SN shock.

While the Fe-K emission line at 6.55 keV indicates that the hard
component is dominated by thermal X-ray emission, a non-thermal
component might be hidden underneath.  To determine an upper limit on
this component, we fit the hard band (4--9 keV) with a power law
model, plus a Gaussian line to account for Fe.  The fit yields a
photon index, $\Gamma$ = 3.4 (2.5--4.5, 3$\sigma$) and a reduced
$\chi^2_{\nu}$=1.0 ($\nu$=143).

We also fitted the entire 0.6--9 keV range of the \asca \ data with a
two-component model: \mekal \ to account for the soft component, and a
power law or a synchrotron model, \srcut, to represent the hard
component.  A Gaussian line was also added near 6.55 keV to account
for Fe-K emission.  The synchrotron model, \srcut, is more appropriate
for describing synchrotron X-ray emission from SNRs than a power
law. It models the particle spectrum cutting off exponentially, and is
parameterized by the radio spectral index and flux density, and a
characteristic roll-off frequency, $\nu_{roll}$ (Reynolds 1998;
Reynolds and Keohane 1999).  This gives the sharpest plausible
roll-off in a synchrotron spectrum, while the power-law model has no
roll-off at all.  A true synchrotron description should lie between
these extremes.  In Table~4, we summarize the parameters of these
fits.
The \mekal+ \power \ law and \mekal+\srcut \ models
yield poorer fits than the \mekal+\shock model ($\chi^2_{\nu}$ =
1.75--1.95), as they do not account for the line emission from Sulfur
or Iron.

We subsequently added the PCA spectrum to test for non-thermal
emission in the 5--15 keV band.  Fitting the \asca \ and the PCA data
in the 0.6-15 keV with a \mekal+ \power \ law model, plus Gaussian
lines to account for the emission from Argon and Fe-K, yields a power
law index $\Gamma$=4.23 (4.15--4.31, 3$\sigma$), and a reduced
chi-squared $\chi^2_{\nu}$=1.79 ($\nu$=763). This model is again worse
than the \mekal+\shock model (with solar abundances), implying that
the data favor the thermal model.

Adding a power law component to the \mekal+\shock model (shown in
Table~3) improves the fit (an F-test yields a probability of 5
$\times$ 10$^{-4}$).  The \mekal+\shock+\power \ law model yields
$kT_s$=1.44 keV and
$n_0t$=8.6$\times$10$^{10}$ (cm$^{-3}$ s) for the hard component.  The
power law component is characterized by a photon index $\Gamma$ $\sim$
1.5, and a flux, $F_x$ (5--15 keV) $\sim$ 1.2 $\times$10$^{-12}$ erg
cm$^{-2}$ s.  However, the complexity of the model used and the high
contamination by the Galactic ridge leave its parameters highly
uncertain. We note that fits to the \asca \ hard band (4-9 keV)
plus the PCA spectrum (5--15 keV) with a \shock\ model (as described
in the previous section) do not require an additional power
law, as the derived shock temperature $kT_s$=2.3 (2.1--3.1) keV is
higher than the temperature derived above ($kT_s$=1.44 keV). In
summary, the \rxte \ data favor a thermal model for the hard
component, and do not allow us to constrain the parameters of an
additional power law component, due to the complexity of the model
used, the large number of fitted parameters, and the uncertainties in
modeling the background.

\section{Timing results}

To search for pulsations in the \asca \ data, we extract the source
events using the GIS detectors in the high-bit rate mode, with the
standard time resolution of 62.5 milliseconds.  We perform power
spectral density (PSD) analysis on the barycenter corrected photon
arrival times and search for pulsations in the 0.01--8 Hz frequency
range. The upper frequency is dictated by the time resolution of the
data. No pulsations were found at a significant level ($\geq$
3$\sigma$).

We computed PSD analyses at two energy bands: soft (0.5--2.4 keV) and
hard (2.5--10 keV).  We bin the data into 0.5 s bins, and perform a
long FFT on the background-subtracted, binned light curves.  This
allowed us to search for pulsations up to 1 Hz.  For the higher
frequency range (up to 8 Hz), we use the 62.5 millisecond time
resolution and performed average FFT's with 512 s length each.

We find some interesting peaks, but no detection was found with a high
confidence level.  We subsequently fold the data at the peaks
determined from the PSD's using a Z$_n^2$ test.  At the soft energies,
no pulsations were found at a level $\geq$ 3$\sigma$, and the pulsed
fraction is $\leq$ 7\% in the 0.01--8 Hz range.  At the higher
energies (2.5--10 keV) range, no pulsations were found with a
confidence level $\geq$ 1$\sigma$, and the pulsed fraction was $\leq$
12\%.

We use the \rxte \ data to search for higher frequency pulsations.
The PCA data observed with Good Xenon modes provide a $\mu$s time
resolution.  We select the events in the 5--20 keV energy range and
from the top layer in order to maximize the signal to noise ratio.  We
apply the barycentric correction, and compute average FFT's to search
for any coherent pulsations up to 128 Hz. No pulsations were found.
The upper limit on the pulsed fraction is $\sim$ 15\% in the 5--20 keV
range (where we have estimated that $\leq$ 10\% of the source count
rate originates from a plerion).

\section{Discussion}

\subsection{The diffuse emission}

In the following, we discuss the origin of the diffuse X-ray emission
in the light of the interaction between the SNR material and the
surroundings. In the standard picture of the X-ray emission from young
SNRs, the soft component arises from shocked ejecta and the hard
component is usually attributed to the blast wave.  For this remnant,
we find that high metal abundances improve the spectral fitting,
suggesting that at least part of its X-ray spectrum may be associated
with the ejecta. While the absolute values of the abundances are model
dependent, we find that the hard component (4--9 keV) could be fitted
with a NEI model (\shock) with solar abundances, and does not require
large metal abundances.  It is therefore reasonable to assume that the
hard component is associated with the blast wave, and the soft
component with the ejecta, most likely of a core-collapse SN as
derived from the observed abundance pattern. It is also possible that
the hard component is not due to material shocked by the blast wave,
but results from the shock entering very low-density regions, and that
we should interpret the soft component as a Sedov blast wave. We
discuss both these possibilities below.

\subsubsection{Young, Ejecta-Dominated Remnant of a Core-Collapse SN}

If 3C~397 is an ejecta-dominated SNR and the hard component associated with
the blast wave, then we may estimate the
parameters of the SN explosion using the Sedov model (Table~1).
We use a distance of 10 kpc to 3C~397, and estimate the physical
parameters in units of $D_{10}$.  From the equation in HSC~83
below Equation (10), we can convert the measured emission measure ($EM$) 
into an upstream density $n_0$.  From the \rosat \ HRI image (DR~99),
 the mean angular radius is about $1\farcm8$, implying
$r_s  = 5.3 D_{10}$ pc.  From Table~1 (\shock + \sedov
non-equipartition fit, last row),
we determine $\int n_e n_H dV = 10^{14} (4 \pi D^2)
(EM) = 8.4 \times 10^{58} D_{10}^2$ cm$^{-3}$; then from HSC~83:
$n_0 (EM) = 5.64 \times 10^{-29} r_s({\rm pc})^{-3/2} \left( \int
{n_e n_H dV} \right)^{1/2} =
1.33 D_{10}^{-1/2} \ {\rm cm}^{-3}.$
This implies a swept-up mass of 29 $M_\odot$ (assuming
a mean mass per particle $\mu$ of 1.4), consistent with a
massive progenitor and an evolutionary stage between ejecta-dominated
and Sedov. 
The parameters of the SN derived using the Sedov non-equipartition
fit to the hard component are summarized in Table~5.

The Sedov spectral fit actually overdetermines the SNR parameters,
since from the observed shock temperature and ionization time-scale
alone we can find the remnant age and upstream density.  Equations (4a
-- 4f) in HSC~83, with our measured values of $T_s = 2.01 \times 10^7$
K and $n_0 t = 9.6 \times 10^{10}$ cm$^{-3}$ s, give $v_s = 1,190$ km
s$^{-1}$, $E_0 = 0.83 \times 10^{51} D_{10}^2$ erg, $t = 1,750 D_{10}$
yr, and $n_0 ({\rm Sedov}) = 1.73 D_{10}^{-1}$ cm$^{-3}$.

This value of $n_0$ is, within the various uncertainties, consistent with
that determined from the measured $EM$. The swept-up
mass from $n_0 ({\rm Sedov})$ is 40 $M_{\odot}$; in either case,
we find a large mass, indicating massive progenitor, since if the
ejected mass were only 1.4 $M_\odot$, the remnant should show little
or no evidence of ejecta by this time.  In general, the morphology of
the remnant, its location in the Galactic plane, and the suggestion
that the soft component dominating the X-ray spectrum is due to
ejecta, favor a core-collapse explosion.

We note that using the fit to the hard component only (4-9 keV), the
\shock model yields a higher temperature $kT_s$=2.45 keV (1.8--4.2
keV), and a slightly lower ionization time-scale $\tau$=$n_0t$= 3.1
(1.5--8.3) $\times$10$^{10}$ cm$^{-3}$ s.  The observed emission
measure of $EM$=0.0224 corresponds to $\int n_e n_H dV = 2.7 \times
10^{58}$ cm$^{-3}$.  If the emission volume $V = f V_{\rm tot}$, with
$f$ the filling factor, we find an upstream density $n_0 = 0.27
f^{-1/2}$ cm$^{-3}$ implying a shock age $t = 4,100 \ (1,600 - 9,500)
\ f^{1/2}$ yr. For a reasonable filling factor of 0.25, these
estimates are quite comparable to those from the Sedov fits above
and strengthen our confidence in them.

We have thus accounted satisfactorily for the gross remnant properties
using only the high-temperature component of the X-ray emission.
What, then, of the low-temperature component (Table~1) with an
emission measure larger by about 600.  If that component represents
shocked ejecta, its mass cannot greatly exceed that of shocked ISM.
The only way the emission measure can be greatly increased is if that
material is very highly concentrated in small regions, since for a
given total mass $M_{\rm ej}$, $\int n_e n_H dV \propto M_{\rm ej}^2
f_{\rm ej}^{-1}$ with $f_{\rm ej}$ the ejecta filling factor.  Our
fits then suggest that the ejected material, which is unlikely to
comprise more than half the swept-up mass of order 30 $M_\odot$, is
concentrated in very small regions.  The \rosat \ HRI image (Figure
12, DR~99), sensitive to the energy range from 0.4 to 2 keV, should
illustrate the spatial location of the soft component material.  That
material appears to be distributed more or less like the radio
emission, largely concentrated near the edges (especially the western
edge), with a small, bright region in the interior.  However, if that
material is concentrated into knots, the image might resemble
what is observed.  A good example of such clumpy ejecta is provided by
optically-emitting O-rich knots in Cas~A, which are the most dense,
undecelerated ejecta fragments plowing through the ambient
circumstellar medium. Because of their high velocities, pressures in
these knots are much higher than in the bulk of the shocked ejecta. We
envision a similar situation in 3C~397, where the soft X-rays with a
large emission measure are produced by dense, fast-moving ejecta
clumps, while emission from the large-scale reverse shock should be
harder and much fainter. The ram-pressure compressed clumps of ejecta
must be at pressures at least an order of magnitude higher than the
pressure of the ambient, much more tenuous X-ray emitting gas, because
of the factor of 600 larger $EM$ for the soft X-ray component (we
expect at most a factor of $\sim 50-100$ larger $EM$ for the X-ray
emission from a large-scale reverse shock). Future high spatial
resolution observations with \chandra \ and \xmm \
should provide enough information to confirm or refute this picture.

\subsubsection{A Medium-Aged SNR in a Dense ISM}

The soft X-ray component may alternatively be identified with the
blast wave. While we obtained poor fits with the \sedov model, this
component is well fit by the \mekal \ model with temperature $kT_l =
0.175$ keV (Table~1). If this temperature is identified with the
post-shock temperature, we obtain 375 km s$^{-1}$ for the blast wave
velocity. The preshock density $n_0$ may be estimated by noting (for
comparable filling factors for the two components) that $n_0 \propto
EM^{1/2}$ and that the emission measure ratio between the low- and
high-temperature component is equal to $\sim 600$. This gives $n_0(EM)
= 33 D_{10}^{-1/2}$ cm $^{-3}$, or a mean postshock electron density
of about 160 cm$^{-3}$, implying that the SN progenitor exploded in a
particularly dense environment. When combined with the remnant's
angular size, we obtain the total swept mass $M_s = 570 D_{10}^{5/2}$
$M_\odot$. Assuming that the remnant's dynamics can be well described
by a Sedov dynamics, we estimate the total SN kinetic energy $E$ at
$1.2 \times 10^{51} D_{10}^{5/2}$ ergs, parameter $\eta = n_0^2E = 1.1
\times 10^{54} D_{10}^{3/2}$ ergs cm$^{-6}$, ionization timescale
$n_0t = 5 \times 10^{12} D_{10}^{3/2}$ cm$^{-3}$ s, and the SNR age $t
= 5300 D_{10}^2 yr$. At this stage of its evolution, the remnant may
be at the transition from the Sedov stage to the radiative stage,
because this transition should occur at $t_{tr} = 2.9 \times 10^4
E_{51}^{4/17} n_0^{-9/17}~{\rm yr} = 5000$ yr, an age equal to the
estimated SNR age.

The high-temperature component in this picture of a middle-aged SNR
must come from the hot interior of 3C~397, occupied by gas shocked by
a high velocity shock earlier in the evolution of the
remnant.
It is likely that
this gas completely fills the remnant's interior, i.~e., its volume
filling
fraction is equal to $\sim 0.75$. With this filling fraction, we
deduce that its electron density $n_e$ is approximately equal to
2.5 cm$^{-3}$. Because ionization
timescale of this hot component is equal to $n_0t = 3.9 \times 10^{10}$
cm$^{-3}$ s (or $n_et = 1.9 \times 10^{11}$ cm$^{-3}$ s) in the \snei
model (Table~1), the hot gas was shocked about $n_et/n_e \sim  2500$ yr
ago,
certainly a reasonable timescale for a 5000 yr old remnant. The mass
of the hot gas is equal to 30 $M_\odot$, and its pressure appears to be
6--8 times lower than pressure in the low-temperature component. This
is somewhat lower than a ratio of 3 expected in a Sedov model. It is
possible
that the volume filling fraction of the high-temperature component is
2--3 lower than we assumed, and then its pressure would be higher than
derived above. If this is the case, then most of the SNR volume would
have to
be occupied by a very tenuous hot gas with a negligible emission
measure.

The origin of the high-temperature component is not clear in the
framework of the middle-aged SNR. Its presence may indicate
significant departures from a uniform ambient medium, because Sedov
models cannot produce such a strong high-temperature
component. Because mass of the hot interior gas is nearly 20 times
smaller than the mass of the swept shell, this material was shocked
early in the evolution of the remnant, and originated relatively close
to the SN progenitor. We expect the SN progenitor to be a massive
star, which is likely to be found near the place of its birth and
associated with dense ISM. But massive SN progenitors modify the
distribution of the ambient medium in its vicinity, blowing stellar
winds and creating dense gaseous shells. We then would naturally
expect deviations from the Sedov dynamics early in the evolution of
the remnant. Perhaps the hot-temperature component is the relic from
earlier stages of the SNR evolution, when the blast wave encountered
an ambient medium strongly modified by the SN progenitor.

Another possibility is that the overall dynamics of 3C~397 are poorly
described by the Sedov solution. We have already concluded that
radiative cooling is likely to be important in the 3C~397 shell, which
could decrease the shell temperature and enhance soft X-ray
emission. The ambient ISM may also be clumpy, a likely possibility in
view of a generally inhomogeneous nature of dense ISM, which could
also affect the remnant's dynamics and its X-ray emission. Finally,
neglected physical processes such as electron thermal conduction may
have caused significant departures from the Sedov solution. For
example, Cox \ea~(1999) estimate that the central density at the time
of transition from the Sedov stage to the radiative stage is
approximately 10 times lower than the preshock density in SNR models
with thermal conduction. Because the density ratio between the
preshock gas and the high-temperature component in 3C~397 is also of
this order, we attempted to fit \asca \ spectra with the thermal
conduction models kindly provided to us by Randall Smith (these models
were used by Shelton \ea \ 1999 to model SNR W44). While the resulting
fits have not produced better results that our one-component fit with
the {\it SEDOV} model, models with thermal conduction are still a
viable alternative because the existing set of models was designed for
SNRs with much lower preshock densities, such as W44.  A more serious
problem might be the lack of evidence for elemental enrichment in the
hot component, as we would expect a moderate enhancement of Fe in the
remnant's interior.  A detailed study of the 3C~397 dynamics, clearly
outside the scope of our present work, is required in order to
understand the nature of the high-temperature component in the
framework of a middle-aged SNR.

\subsection{A hard non-thermal tail?}

In the following, we discuss the possibility that the hard component
is synchrotron emission from highly relativistic particles accelerated
at the SNR shock.  We show that a power-law description results in
reasonable parameter values, while a somewhat better motivated cutoff
synchrotron model (Reynolds 1998) describes the data as well, giving
results consistent with Reynolds and Keohane (1999) which involved
highly simplified spectral fits.  For power-law models, the index,
$\Gamma$, is highly dependent on fitting the soft component, being
steep ($\Gamma$ $\sim$ 3.4) when the soft component is fitted with a
\mekal \ or \shock \ model, and harder for a Sedov fit ($\Gamma$
$\sim$ 2).  In the following, we estimate the equipartition magnetic
field and the non-thermal energies, whose values are not too sensitive
to the power law photon index.  Using the power law fit parameters to
the hard component only (4--9 keV), the photon index $\Gamma$ is 3.4,
and the flux $F_x$ (4--9 keV) is 3.8 $\times$ 10$^{-12}$ erg cm$^{-2}$
s$^{-1}$, corresponding to a luminosity $L_x$ (4--9 keV) of 4.5
$\times$ 10$^{34}$ $D_{10}^2$ erg s$^{-1}$ (unabsorbed).  We use the
\asca \ hard band only since the PCA spectral fitting is highly
dependent on the emission from the Galactic ridge.  Assuming
equipartition between energy in the relativistic electrons and
magnetic fields, we estimate a magnetic field $B$ $\sim$
1.2$\times$10$^{-5}$ ($f_{s} \theta_{1.8}^3 D_{10}) ^{-2/7}$ G, where
$f_{s}$ is the fraction of the flux that is synchrotron radiation.
The total energy in the electron distribution out to X-ray emitting
energies is $U_e$ $\sim$ 0.7 $\times$ 10$^{45}$ $B_{-5}^2$ erg;
$B_{-5}$ being the magnetic field in units of 10$^{-5}$ G.  This
represents a small fraction of the SN explosion energy.  The
synchrotron lifetime of an electron emitting $\sim$ 9 keV X-rays is
$\tau_{1/2}$ $\sim$ 650 $B_{-5}^{-3/2}$ years.  The typical electron
energies producing the synchrotron photons of energy $E_{\gamma}$ can
be estimated to be $E_e \ \sim \ 150 \
B_{-5}^{-1/2} \left(\frac{E_{\gamma}}
{9 \ {\rm keV}}\right)^{1/2}$ TeV.  

A power-law spectrum is not expected to arise naturally in the
high-energy part of the electron spectrum; rather, one expects a slow
rolloff from the low-frequency synchrotron power-law.  The simplest
description of this rolloff is the synchrotron spectrum from an
exponentially cut off power-law electron distribution $N(E) = K E^{-s}
\exp(-E/E_{max})$, the ``cutoff'' synchrotron model (Reynolds 1998),
called $SRCUT$ in XSPEC.  In $SRCUT$, the fitted roll-off frequency,
$\nu_{roll}$, is related to the cutoff electron energy, $E_{max}$, via
the relation: $\nu_{roll}$ $\sim$ 0.5 $\times$ 10$^{16}$
$\frac{B}{10^{-5}G}$ $\left(\frac{E_{max}}{10 \ TeV}\right)^2$ Hz.
Using the fitted value of $\nu_{roll}$ = 2.89 $\times$ 10$^{16}$ Hz,
and an equipartition magnetic field of 10$\mu$G, we estimate a cutoff
electron energy of 24 $B_{-5}^{-1/2}$ TeV.  This value is in agreement
with the upper limit derived by Keohane (1998) and Reynolds and
Keohane (1999).  A synchrotron explanation for the hard component
results in reasonable parameters whether it is described by a power
law or by the cutoff model; better observations will be needed to see
if such a component is demanded by the data.

\subsection{A hidden plerion?}

The presence of a pulsar-powered component (plerion) is suggested by
the HRI image showing the hot spot at the center of the remnant.  We
subsequently estimate the intrinsic parameters from a hidden pulsar
taking the hard component luminosity as an upper limit on a plerionic
contribution.  We assume a Crab-like plerion, and use the empirical
formula derived by Seward \& Wang (1988), $logL_x$ (erg s$^{-1}$) =
1.39 $log{\dot{E}}$ -16.6;
where $L_x$ represents the X-ray luminosity of the plerion in the 0.2--4 keV
band.  The power law model for the hard component implies an observed flux
$F_x$ (0.2--4 keV) = 6.2 $\times$ 10$^{-12}$ erg cm$^{-2}$ s$^{-1}$;
which translates to a luminosity $L_x$ (0.2--4 keV) =
4.7 $\times$ 10$^{36}$ erg s$^{-1}$ (after correction for absorption).
This implies a spin-down luminosity of $\dot{E}$ $\leq$ 2 $\times$
10$^{38}$ erg s$^{-1}$, and a period of
$P$ $\geq$ 0.08(s) ($t_3$ $\dot{E}_{38}$)$^{-1/2}$; where
$\dot{E}_{38}$ is the spin-down luminosity in units of
10$^{38}$ erg s$^{-1}$, and $t_3$
is the pulsar's age in units of 10$^3$ years.
We stress that these parameters are derived assuming that 
the hard component arises entirely from a plerion.
These are extreme limits since the hard band GIS image 
indicates that the hard
X-ray emission peaks at the western lobe, with
some emission from the central spot (Figure 2).
Assuming that only $\sim$ 10\%
of the hard component arises from
the central hot spot, then the spin-down luminosity would
be $\leq$ 4$\times$10$^{37}$ erg s$^{-1}$, and the
period $P$ $\geq$ 0.12 $t_3$$^{-1/2}$ s.

A synchrotron component in the X-ray spectrum of 3C~397 could be also
associated with a central engine injecting relativistic particles,
which as they encounter a strong shock, would produce the high-energy
non-thermal tail (up to $\geq$ 9 keV). Such a process is known to
occur in the SNR W50 powered by the jet source SS433 (Safi-Harb \&
Petre 1999 and references therein).  The axial ratio of 2:1 in 3C~397
is similar to that in W50, in which the bloating results from the
interaction between the jets in SS433 and the SNR shell.

\section{Summary and Conclusions:}

We have presented \rosat,  \asca, \ and \rxte \ observations of 3C~397.
The \rosat \ high-resolution HRI image shows a central hot spot, possibly associated
with a compact object whose nature remains a mystery, as no X-ray pulsations
nor a radio counterpart have been found.
The \asca \ and \rosat \ images show that the
remnant is highly asymmetric, having a double-lobed
morphology similar to the radio shell.
The hard band image obtained with the \asca \ GIS
overlayed on the \rosat \ HRI image shows that 
the hard emission peaks at the western lobe,
with little hard X-ray emission originating from the central spot.

The spectrum is heavily absorbed, and dominated by thermal emission
with emission lines evident from Mg, Si, S, Ar and Fe.
Single-component models fail to fit the \asca \ spectra (0.6-9 keV).
Even a \sedov model (a NEI model including a range of temperatures) does
not account for the emission above $\sim$ 4 keV.  Two components, at
least, are required to fit the data: a soft component, characterized by
a large ionization time-scale, and a hard component, required to
account for the Fe-K emission line and characterized by a much lower
ionization time-scale.  We use a set of NEI models, and find that the
fitted parameters are robust. The temperatures from the soft and hard
component are $\sim$ 0.2 keV and $\sim$ 1.6 keV respectively.  The
corresponding ionization time-scales $n_0 t$ are $\sim$ 6 $\times$
10$^{12}$ cm$^{-3}$ s and $\sim$ 6 $\times$ 10$^{10}$ cm$^{-3}$ s respectively.
The large $n_0 t$ of the soft component indicates that it is approaching
ionization equilibrium, and it can be fitted equally well with a
collisional equilibrium ionization model.  The 5--15 keV PCA spectrum,
though contaminated by the emission from the Galactic ridge, allowed us
to confirm the thermal nature of the hard X-ray emission.  Fitting the
hard component (5--15 keV band) yields, however, a higher shock
temperature ($kT_s$ $\sim$ 2.3 keV) than the one derived from fitting
the entire band with two-component models (Table~1).  A third component
originating from a pulsar-driven component is possible, but the
contamination of the source signal by the Galactic ridge did not allow
us to determine its parameters, or find pulsations from any hidden
pulsar.

We discuss the two-component model in the light of two scenarios:  a
young ejecta-dominated remnant of a core-collapse SN, and a medium-aged
SNR in a dense ISM.\\
In the first scenario, the hot component would
arise from the blast wave and the soft component from the ejecta.
The derived age (a few thousand years) and the presence
of a central X-ray source makes 3C~397 similar to the 
young SNRs G11.2--0.3 (Vasisht
\ea \ 1996), Kes~73 (Gotthelf \& Vasisht 1997), and RCW~103 (Petre \&
Gotthelf 1998). G11.2--0.3 harbors a hard X-ray plerion powered by a fast
millisecond pulsar (Torii \ea \ 1997). Kes~73 and RCW~103 harbor
radio-quiet X-ray sources:
an anomalous X-ray pulsar in Kes~73 (Vasisht \& Gotthelf 1997), and a
low-mass X-ray binary candidate in RCW~103 (Garmire \ea \ 2000,
IAU Circ 7350). 
If a central neutron star exists in 3C~397, as suggested by the \rosat
\ HRI image, then it must be radio quiet.
The absence of both a radio
spot and X-ray pulsations could be then attributed to a
cooling neutron star, an anomalous or binary X-ray pulsar,
or a weak X-ray plerion buried underneath the SNR.
\rosat \  did not allow an accurate measurement
of its spectrum due to the heavy 
interstellar absorption towards 3C~397,
and the narrow energy band of \rosat.
Future \chandra \ observations will unveil its nature.
We note that recently, \chandra \ observations of CasA SNR
revealed a central radio-quiet X-ray source
(Pavlov \ea \ 2000),
 and that a new 424 ms (radio-quiet) X-ray pulsar
has been discovered with \chandra \ observations of
the SNR PKS 1209-52 (Zavlin \ea \ 2000).
The hybrid X-ray morphology of 3C~397, with both shell and central emission, combined
with its age, makes it a unique SNR--perhaps a transition object from a
shell (like the historical SNRs) into a composite that is well into the
Sedov phase of evolution (like Vela).

In the second scenario (a middle-aged SNR), the soft component would
represent the emission from the SNR expanding in a dense medium.  The
hard component would arise from the hot interior shocked by a fast
shock earlier in the evolution of the remnant. Alternatively, a Sedov
model invoking thermal conduction would modify the Sedov dynamics, and
produce a hot inner component (as was proposed for the SNR W44; Cox \ea
~1999).  In this scenario, the SNR would be entering its radiative
phase, and would emit in the infrared. 
A 1ks exposure obtained with the
two-micron All-Sky Survey (\2mass) Infrared telescope at IPAC reveals,
in the J band, faint diffuse emission in the northern part of the SNR
(J.~Rho, private communication).  The extent of the diffuse emission to
the west is, however, confused with bright stars in the field.  
No emission was detected from 3C~397 in the far infra-red.
Furthermore, HI and CO observations should reveal the presence of a
radiative shell and its interaction with a dense medium (as found for
W44, IC 443, and 3C391: Chevalier 1999).  To our knowledge, no evidence
of an HI shell or CO emission is found to be associated with 3C~397.  The
current picture we present here is therefore marginally consistent with
this scenario, but it can not be excluded.

Differentiating between the two scenarios requires a spatially resolved
spectroscopy as well as more detailed modeling (which is outside the
scope of this paper).  In particular, high spatial resolution X-ray
data combined with broadband energy coverage is required to resolve the
central component from the outer shell, and unveil the nature of the
mysterious X-ray spot. This could be achieved with \chandra \ and
\xmm.  Furthermore, future gamma-ray observations with high-spatial
resolution (such as $\it{GLAST}$) will help look for a high-energy tail
and test for models with a radiative shell.  Since 3C~397 is heavily
absorbed, infra-red observations will be crucial to trace the presence
of radiative shocks.  In particular, searching for [OI] 63$\mu$m line
emission will be a powerful probe to test for the radiative model.
Finally, observations with millimeter telescopes will enable us to get
a better estimate of the kinematic distance to 3C~397, and measure the
density profile of the medium into which it is propagating.  We have
observed 3C~397 with the \sest \ telescope (Chile), and the data are in
the process of being analyzed (Durouchoux \ea \, in preparation).

\noindent{\bf Acknowledgments} \\ We greatly acknowledge useful
discussions with A. Valinia on the X-ray emission from the Galactic
ridge, and thank her for providing us with the \rxte \ scans of the
ridge.  We particularly thank E. Gotthelf for his help in using the
ftool {$\it ascaexpo$}; U. Hwang \& G. Allen for scientific
discussions. We are grateful to R. Smith for providing us
with his thermal conduction code,
and J. Rho for her input on the \2mass \ images,
prior to their publication.
We thank the referee, Don Cox, for his careful reading and
invaluable comments and suggestions.
 \\ This research made use of data obtained with the High
Energy Astrophysics Science Archive Research Center (HEASARC) Online
Service and the NASA's Astrophysics Data System Abstract Service
(ADS), provided by the NASA/Goddard Space Flight Center.
S.S.H. acknowledges support from the National Research Council.

\begin{table}[h]

\caption{Two-component model fits to the SIS and GIS spectra.}

\begin{tabular}{c|cc|c}
\hline

Model$^a$ & Soft & Hard & $\chi^2_{\nu}$ ($\nu$) \\
\hline

\shock + \shock  & $kT_l$ = 0.19 (0.185--0.2) keV & $kT_h$ = 1.52 (1.4--1.62) keV
&
1.37 (729)\\
$n_0 t$$^b$ (cm$^{-3}$ s)         & $\tau_l$ = 5.5 $\times$ 10$^{12}$
 &                   $\tau_h$ = 5.9 (4.1--9.8) $\times$10$^{10}$  & -- \\
$EM$$^c$            & 31.5 (24--42) & 0.053 (0.048--0.057) & -- \\
$L_x$$^d$ & 4.8$\times$10$^{38}$ & 2.65$\times$10$^{36}$ & -- \\
\hline

\mekal + \snei & $kT_l$ = 0.175 (0.17--0.18) keV & $kT_h$ = 1.5 (1.34--1.60) keV &
1.57 (730) \\
$n_0 t$ (cm$^{-3}$ s)       &    --          & $\tau_h$ = 3.9 (3.1--5.5)
$\times$10$^{10}$  & -- \\
$EM$            & 43.5 (33--57) & 0.057 (0.050--0.068) & --\\
$L_x$ & 5.5$\times$10$^{38}$ & 1.58$\times$10$^{36}$ & -- \\
\hline

\mekal + \shock & $kT_l$ = 0.175 (0.172--0.18) keV & $kT_h$ = 1.5 (1.39--1.61) keV
& 1.6
(730) \\
$n_0 t$ (cm$^{-3}$ s)        &    --             & $\tau_h$ = 8.5 (5.7--16)
$\times$10$^{10}$  & -- \\
$EM$       &  43.5 (33--56)$^b$ & 0.057 (0.048-0.068) & --\\
$L_x$ & 5.5$\times$10$^{38}$ & 2.2$\times$10$^{36}$ & -- \\
\hline

\mekal + \sedov ($T_e$ = $T_i$) & $kT_l$ =  0.174 (0.17--0.18) keV &
 $kT_h$ = 1.14 (1.0--1.29) keV & 1.66 (730) \\
$n_0 t$  (cm$^{-3}$ s)     & -- &
     $\tau_h$ = 1.1 (0.8--2.0) $\times$10$^{11} $  & -- \\
$EM$            &  43.5 (30--56) & 0.063 (0.053--0.083) & -- \\
$L_x$ & 5.5$\times$10$^{38}$ &  1.21$\times$10$^{37}$ & -- \\

\hline

\mekal + \sedov ($T_e$ $\neq$ $T_i$) & $kT_l$ = 0.177 keV
 & $kT_s$ = 1.735 (1.65--1.9) keV & 1.65 (730) \\
                                   &    & $kT_e$ = 0.87 (0.8--0.9) keV & -- \\
$n_0 t$  (cm$^{-3}$ s)     & -- & $\tau_h$ = 1.17 (0.8--1.85)
 $\times$10$^{11}$ & -- \\
$EM$ & 39  & 0.065 (0.058--0.068) & -- \\
$L_x$ & 4.9$\times$10$^{38}$ &  5.9$\times$10$^{36}$ & -- \\
\hline

\shock + \sedov ($T_e$ $\neq$ $T_i$) & $kT_l$ = 0.187 keV &  $kT_s$ =
1.73 (1.54--1.91) keV & 1.39 (729) \\
  & & $kT_e$ = 0.86 (0.79--0.93) keV & -- \\
 $n_0 t$  (cm$^{-3}$ s)     & $\tau_l$ = 1$\times$10$^{13}$ &
   $\tau_h$ = 9.6 (7.4--11.8) $\times$10$^{10}$ & -- \\
$EM$ & 39 &  0.070 (0.066--0.076) & -- \\ 
$L_x$ & 4.8$\times$10$^{38}$  & 9.8$\times$10$^{36}$  & -- \\
\hline

\end{tabular}

$^a$   The NEI models used are \snei
(constant-temperature single-ionization timescale NEI model),
\shock (plane-parallel NEI shock model), and the \sedov
models  (Borkowski \ea, in preparation).
The subscripts $l$ and $h$ refer to the
low-energy and high-energy components respectively.\\
$^b$ Ionization time-scale;  $n_0$ is the preshock hydrogen density\\
$^c$ The emission measure in units of
$\frac{10^{-14}}{4\pi D^2}$ $\int (n_e n_H dV) (cm^{-5})$. \\
$^d$ The X-ray luminosity in the 0.5--9 keV range (at a distance of 10 kpc).\\
\end{table}

\begin{table}[h]

\caption{PCA observation segments of 3C~397.}

\begin{tabular}{cccc} \\ \hline
 Observation Number & Date & Time  ($\times$10$^4$ s) & PCA$_{L7}$
(\cps)$^a$ \\ \hline
 01-00 & 1997-12-03 & 1.034 & 12.90 $\pm$ 0.07 \\
 01-01 & 1997-12-04 & 1.216 & 13.74 $\pm$ 0.06 \\
 01-02 & 1997-12-06 & 0.488 & 14.97 $\pm$ 0.10 \\
 01-03 & 1997-12-07 & 1.240 & 13.27 $\pm$ 0.06 \\
 01-04 & 1997-12-08 & 0.952 & 12.70 $\pm$ 0.07 \\
 01-05 & 1997-12-08 & 0.910 & 12.85 $\pm$ 0.07 \\
\hline
\end{tabular}

$^a$ Background-subtracted count rates in the 5--15 keV range,  using the L7/240
background model. The uncertainties reflect the statistical errors
only, and do not include any systematic errors associated with the
background subtraction.
\\
\end{table}

\begin{table}[h]
\caption{The \asca \ and PCA data fitted in the 0.6-15 keV range.
The model consists of a \mekal component
to account for the soft component, and a \shock
component to model the hard emission.
The Galactic ridge was fitted with the two-component
model of VM~98}

\begin{tabular}{ccc}\\
\hline \hline
Model & Soft & Hard  \\ \hline
\mekal+\shock & $kT_l$ = 0.175 & $kT_h$ =  (1.46-1.54) keV
\\
  Ionization time-scale   &   --   & $n_0t$ =  (5.3--9.6) $\times$ 10$^{10}$
 (cm$^{-3}$ s)  \\
  EM$^a$          & 43             & 0.058 (0.054--0.061)  \\ \hline
Galactic ridge & $kT_{RS}^a$=3.2 keV & $\Gamma_{power \ law}^a$=1.7  \\
               &  $Norm_{RS}^b$ = 1.315$\times$10$^{-2}$
  & $Norm_{power \ law}^c$ = 7.227 $\times$ 10$^{-3}$
        \\ \hline
\hline
\end{tabular}

$^a$ in units of $\frac{10^{-14}}{4\pi D^2}$ $\int (n_e n_H dV) (cm^{-5})$\\
$^b$ Frozen \\
$^c$ in units of ph cm$^{-2}$ s$^{-1}$ keV$^{-1}$ at 1keV\\
\end{table}

\begin{table}[h]
\caption{The \asca \ data fitted with a \mekal model (soft)
and a nonthermal component (hard) -- a power law or \srcut. A Gaussian line
was added near 6.55 keV to account for the emission from Fe-K.}

\begin{tabular}{ccc}\\

\hline\hline
$kT_l$ (\mekal) & Non-thermal Model & $\chi^2_{\nu}$($\nu$) \\
\hline
0.18 keV & $\Gamma^a$=4.3 & 1.75 (729)  \\
 0.21 keV & $\nu_{roll}^b$=2.9$\times$10$^{16}$ (Hz) & 1.95 (730) \\
            \hline\hline
\end{tabular}

$^a$ Power-law photon index \\
$^b$ Roll-off frequency using the \srcut model \\
\end{table}

\begin{table}[h]
\caption{The parameters of the SN explosion derived from using the Sedov
non-equipartition ($T_e$ $\neq$ $T_i$) fit to the hard component.
The parameters were derived starting with the fitted $EM$ value.}

\begin{tabular}{c|c}\\

\hline

Shock temperature, $kT_s$ (keV) & 1.54--1.91 \\
Ionization time-scale, $n_0t$ (cm$^{-3}$ s)
 & (1.5 -- 2.4) $\times$10$^{10}$ \\
Shock velocity, $v_s$ (km s$^{-1}$) & 1,140--1,270 \\
Age, $t$ (yrs) & 1,800 -- 2,800 \\
Ambient density, $n_0$ & 1.29 -- 1.38 \\
Explosion energy, $E_0$ (ergs) & $(0.62 - 1.0) \times$10$^{51}$ \\
\hline
\end{tabular}

\end{table}

\vfill\eject

\begin{figure}
\centerline{ {\hfil\hfil
\psfig{figure=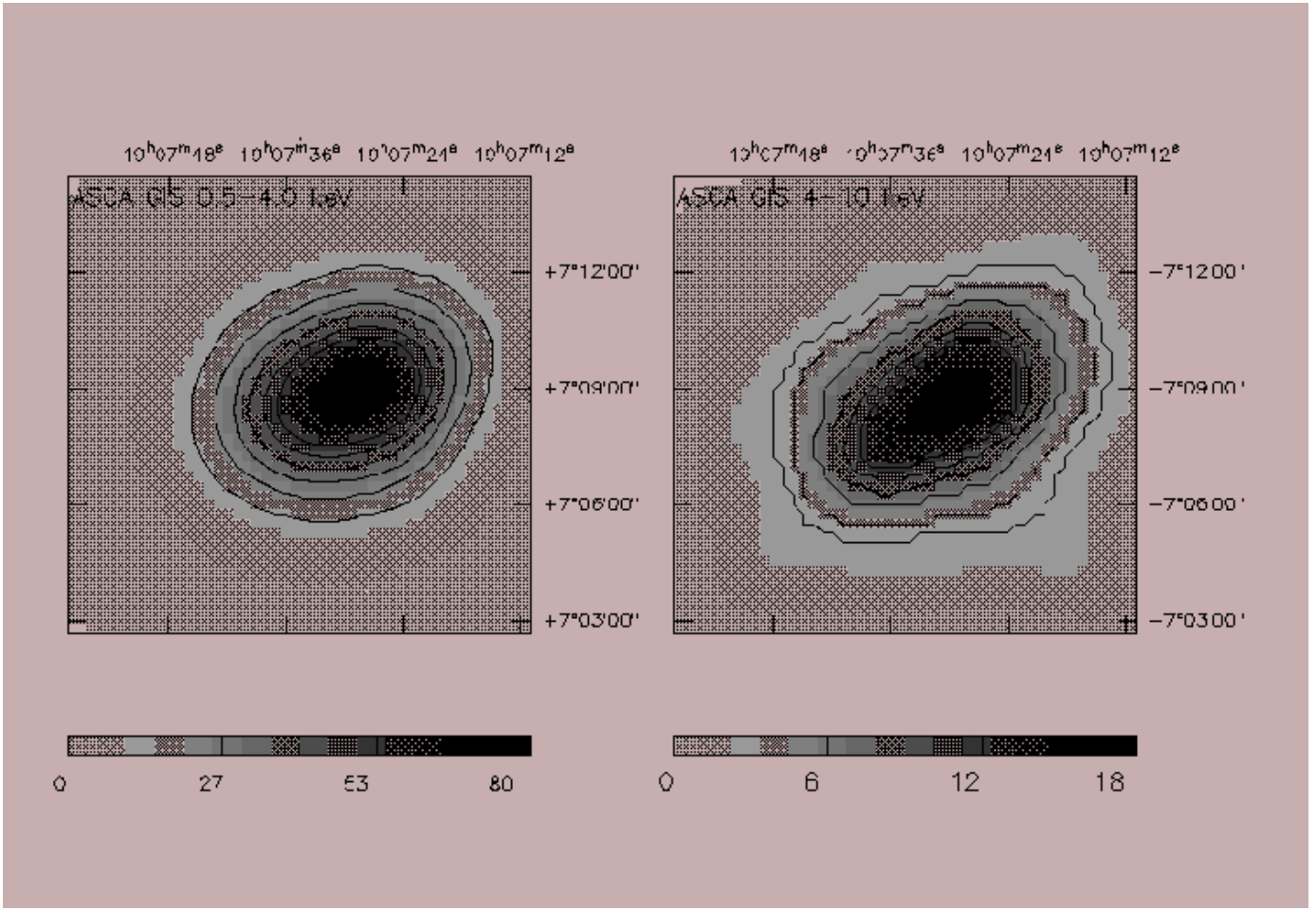,height=5.0in,angle=270}
\hfil\hfil} }
\caption{The \asca \ GIS images of 3C~397 in the soft (0.5-4.0 keV,
left panel) and hard (4.0--10 keV, right panel) energy bands.  The
images are 12\am \ $\times$ 12\am, centered at the
bright hot spot seen with \rosat \ HRI, corrected for vignetting and
exposure, and smoothed with a Gaussian of $\sigma$ = 45\farcs.  Contours
are overlayed on the images with a linear scale.}
\end{figure}

\begin{figure}
\centerline{ {\hfil\hfil
\psfig{figure=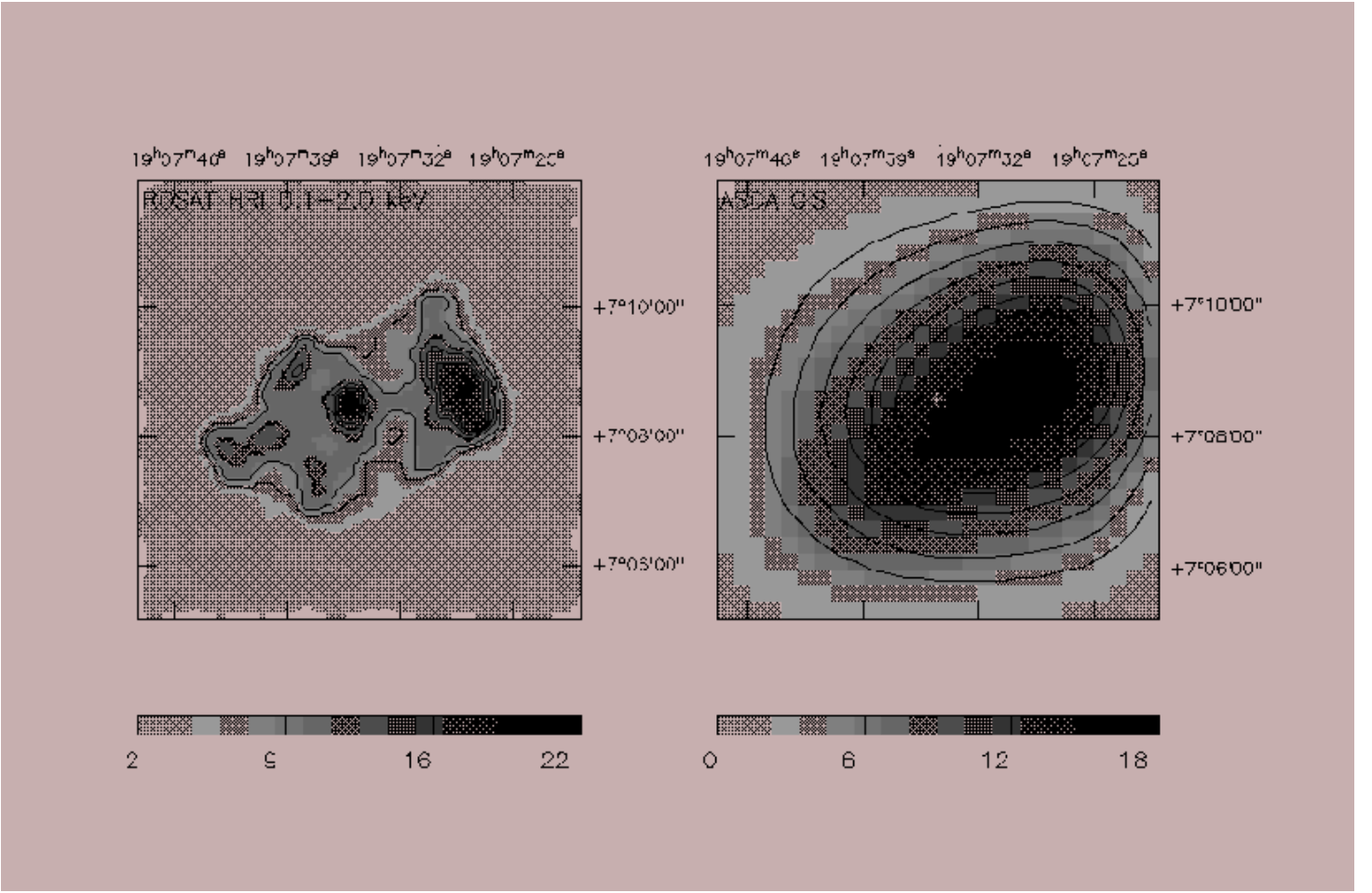,height=5.0in,angle=270}
\hfil\hfil} }
\caption{ 
The \rosat \ HRI image of 3C~397 (left panel) shown with
the \asca \ GIS image at the hard enery band (4--10 keV; right panel).
Both images are 6\farcm8 \ $\times$ 6\farcm8 \, centered at the bright
hot spot shown in the HRI image, and smoothed with a Gaussian of
$\sigma$ = 8\as (45\arcsec) for the HRI (GIS) image.  Contours are
overlaid on the images with a linear scale.  For the GIS hard band
image, we denote the position of the bright spot by a cross, and
overlay contours from the soft energy band (0.5--4 keV) in order to
show that 1) the X-ray emission is not peaked at the bright spot and
2) the hard X-ray emission is more elongated than the emission at the
soft energies.}
\end{figure}

\begin{figure}
\centerline{ {\hfil\hfil
\psfig{figure=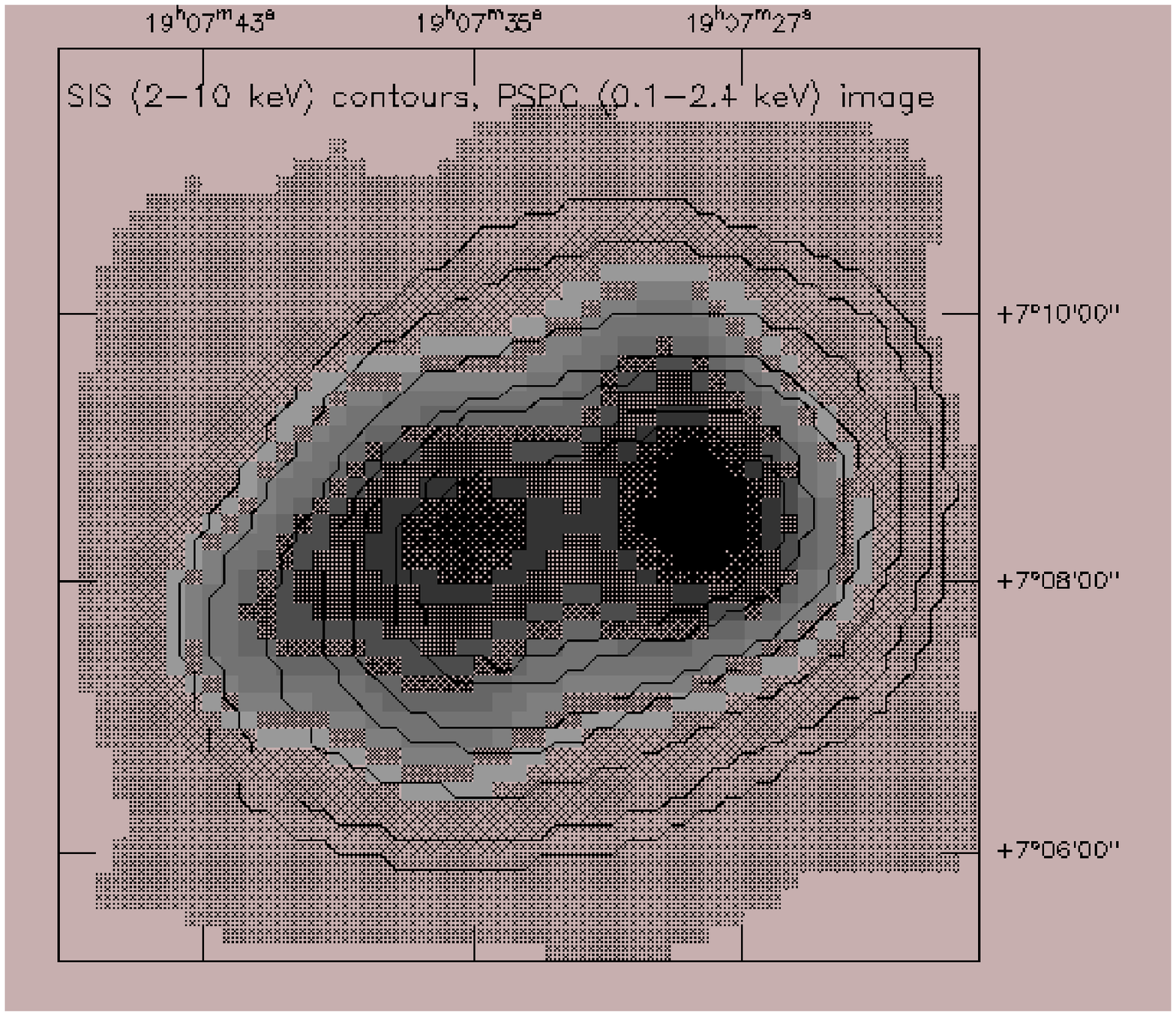,height=5.0in,angle=0}
\hfil\hfil} }
\caption{
The \rosat \ PSPC (0.1--2.4 keV) image  displayed with
the SIS contours corresponding to the hard band (2--10 keV).
The image is smoothed with $\sigma$ = 45\as \ (SIS), and 30\as
\ (PSPC).}
\end{figure}

\begin{figure}
\centerline{ {\hfil\hfil
\psfig{figure=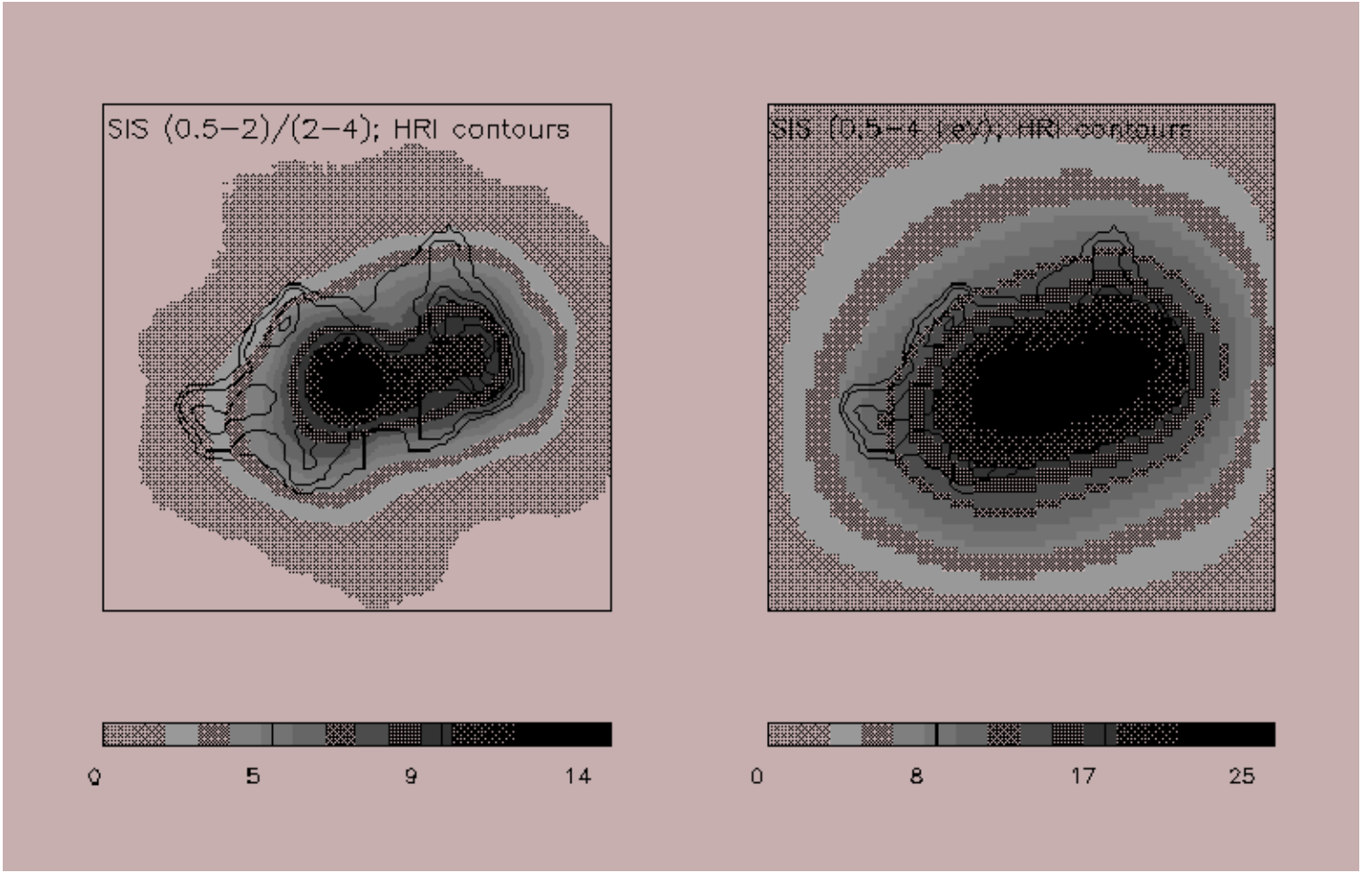,height=5.0in,angle=270}
\hfil\hfil} }
\caption{
The softness ratio
map (left) and the soft (0.5--4 keV) image (right) obtained with  the
SIS.  The images are 6\am.8$\times$6\am.8, centered
at $\alpha$ = 19$^h$ 07$^m$ 34$^s$.6, $\delta$ = 07$^o$ 08$^{\prime}$ 34$^{\prime\prime}$.7
 (J2000).
Contours are overlaid from the \rosat \ HRI image (0.1--2.0 keV),
with a linear scale.  The softness ratio is defined as:
(0.5--2 keV)/(2--4 keV).  }
\end{figure}

\begin{figure}
\centerline{ {\hfil\hfil
\psfig{figure=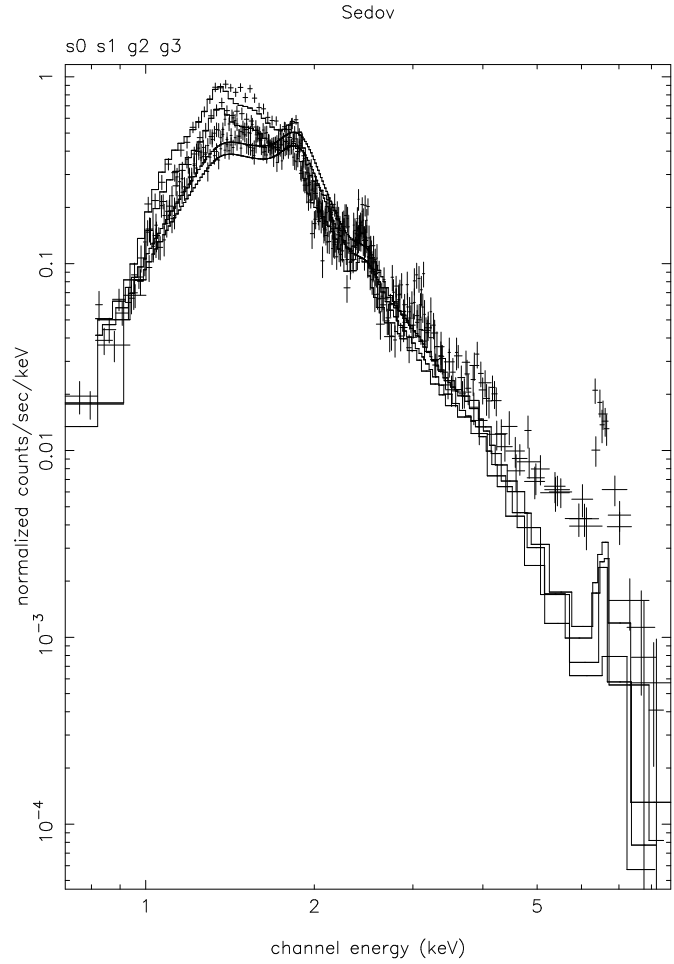,height=5.0in,angle=0}
\hfil\hfil} }
\caption{The \asca \ data (SIS0, SIS1, GIS2, and GIS3: crosses)
fitted with a {$\it SEDOV$}
 equipartition ($T_e$ =
$T_i$) model. The corresponding temperature is $kT_s$ = 0.15 keV, and
the ionization time-scale is $n_0t$ = 1.86 $\times$ 10$^{12}$ s
cm$^{-3}$. The various lines represent the folded model for each instrument.}
\end{figure}

\begin{figure}
\centerline{ {\hfil\hfil
\psfig{figure=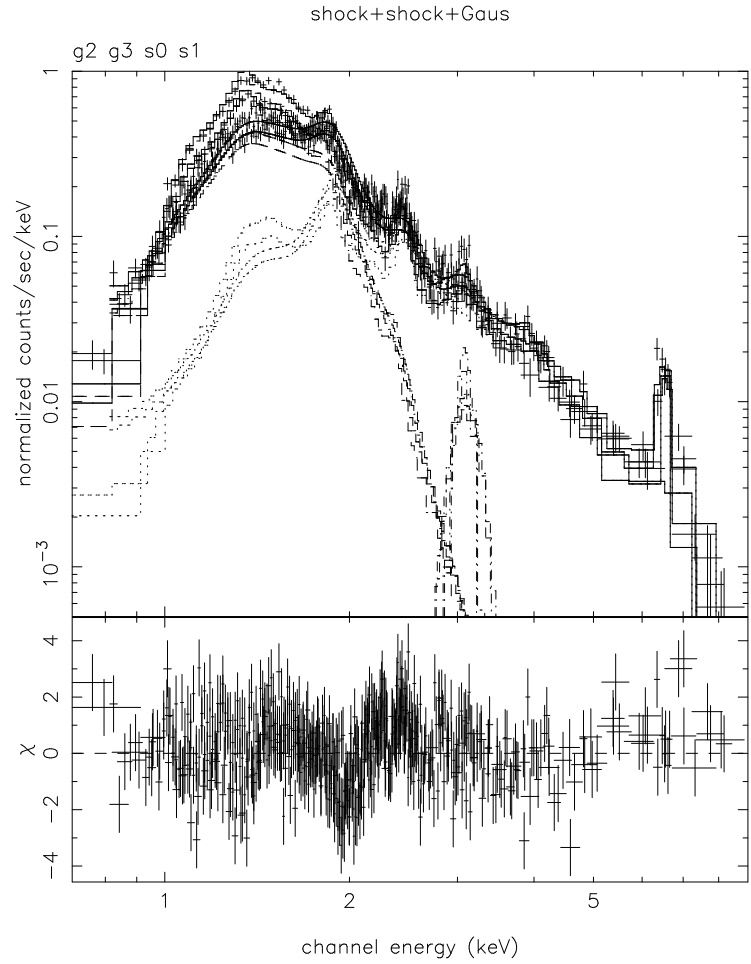,height=5.0in,angle=0}
\hfil\hfil} }
\caption{
The \asca \ data (SIS0, SIS1, GIS2, and GIS3)
fitted with a two-component \shock+\shock
model (Table~1, 1st column). A Gaussian line has been added near 3.1 keV to account
for the emission from Argon, which was not included in the
\shock model. The various lines represent the folded model
for each instrument.}
\end{figure}

\begin{figure}
\centerline{ {\hfil\hfil
\psfig{figure=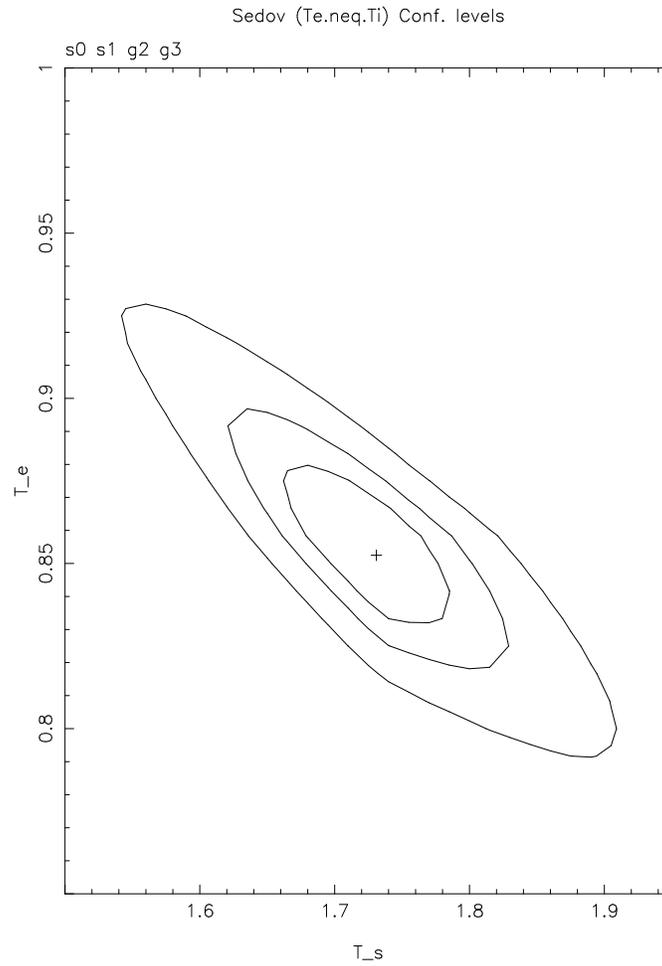,height=5.0in,angle=0}
\hfil\hfil} }
\caption{
The confidence contours for fitting the hard component
with a \sedov model, assuming non-equipartition.
The confidence levels represent
 1$\sigma$, 2$\sigma$, and 3$\sigma$ ranges for the
electron temperature, $T_e$ (keV), and the shock temperature, $T_s$ (keV).
The soft component was fitted with a \shock model.}
\end{figure}

\begin{figure}
\centerline{ {\hfil\hfil
\psfig{figure=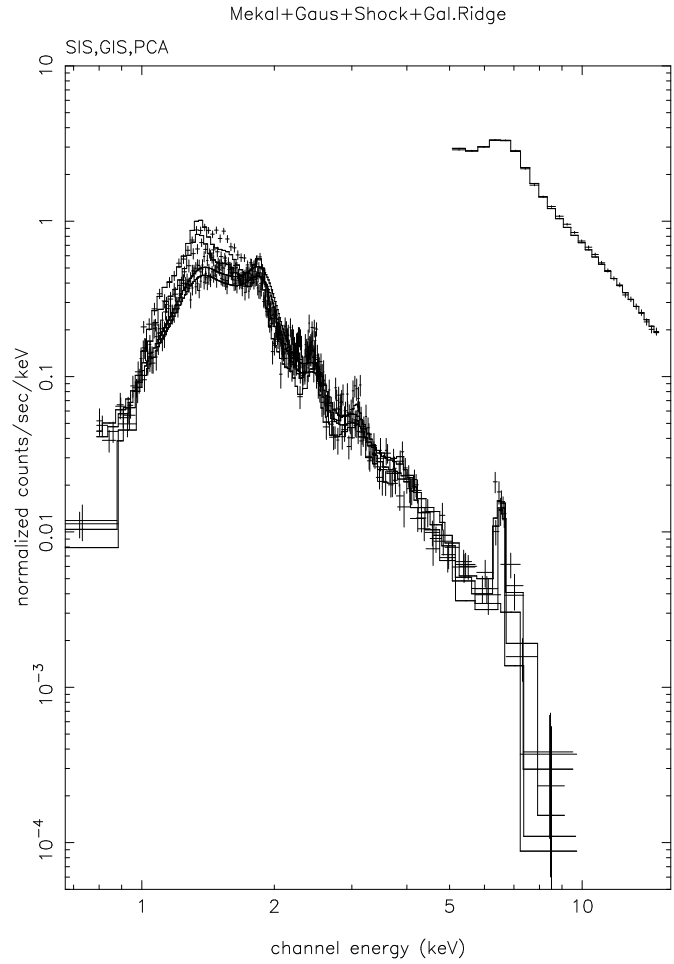,height=5.0in,angle=0}
\hfil\hfil} }
\caption{The \asca \ and \rxte \ PCA spectra fitted
with the two-component (\mekal+\shock model).
The various lines represent the folded model for each instument.
 The Galactic ridge was
fitted with the two-component model of VM~98.  A broad Gaussian line
has been added to the PCA spectrum to account for the Fe line
feature. }
\end{figure}

\end{document}